\documentclass[
journal=jctcce, 
manuscript=article]{achemso}

\usepackage[version=3]{mhchem} 
\usepackage{natbib}
\usepackage{graphicx}
\usepackage{subcaption}
\usepackage{longtable} 
\usepackage{notes2bib}
\usepackage{url}
\usepackage{array}
\usepackage{hyperref}
\usepackage{multirow}
\usepackage{marginnote}
\usepackage{bm}
\usepackage{physics}
\usepackage{braket}
\usepackage{color}

\title{Quantum Embedding Theory for Strongly-correlated States in Materials}

\author{He Ma}
\affiliation
{Department of Chemistry, University of Chicago, Chicago, IL 60637, USA.}

\author{Nan Sheng}
\affiliation
{Department of Chemistry, University of Chicago, Chicago, IL 60637, USA.}

\author{Marco Govoni}
\affiliation
{Pritzker School of Molecular Engineering, University of Chicago, Chicago, IL 60637, USA.}
\alsoaffiliation
{Materials Science Division and Center for Molecular Engineering, Argonne National Laboratory, Lemont, IL 60439, USA.}
\email{mgovoni@anl.gov}

\author{Giulia Galli}
\affiliation
{Pritzker School of Molecular Engineering, University of Chicago, Chicago, IL 60637, USA.}
\alsoaffiliation
{Department of Chemistry, University of Chicago, Chicago, IL 60637, USA.}
\alsoaffiliation[MSD]
{Materials Science Division and Center for Molecular Engineering, Argonne National Laboratory, Lemont, IL 60439, USA.}
\email{gagalli@uchicago.edu}

\begin{document}

\begin{abstract}
Quantum embedding theories are promising approaches to investigate strongly-correlated electronic states of active regions of large-scale molecular or condensed systems. Notable examples are spin defects in semiconductors and insulators. We present a detailed derivation of a quantum embedding theory recently introduced, which is based on the definition of effective Hamiltonians. The effect of the environment on a chosen active space is accounted for through screened Coulomb interactions evaluated using density functional theory. Importantly, the random phase approximation is not required and the evaluation of virtual electronic orbitals is circumvented with algorithms previously developed in the context of calculations based on many-body perturbation theory. In addition, we generalize the quantum embedding theory to active spaces composed of orbitals that are not eigenstates of Kohn-Sham Hamiltonians. Finally, we report results for spin defects in semiconductors.

\end{abstract}

\section{Introduction}
Atomistic, quantum mechanical simulations are playing an increasingly important role in designing functional materials. In the past three decades, density functional theory (DFT) has become a standard method for quantum mechanical simulations of molecules and condensed systems. While DFT has been successfully applied to predict structural and electronic properties of a variety of systems at zero and finite temperature, DFT calculations do not accurately describe strongly-correlated electronic states, i.e. states that cannot be represented by a single determinant of one-electron orbitals \cite{Cohen2008,Su2018}. Prominent examples of strongly-correlated systems include transition metal oxides with localized \textit{d} or \textit{f} bands\cite{Anisimov1997}, deep centers in semiconductors \cite{Bockstedte2018}, and reactions centers in enzymes \cite{Kurashige2013,Sharma2014}. Developing efficient and accurate methods to simulate strongly-correlated electronic states is a long-standing challenge for the electronic structure community.

In principle, any electronic state, whether strongly or weakly correlated, can be described by performing full configuration-interaction (FCI) calculations, which exactly solve the many-body Schr\"odinger equation of electrons subject to the potential of the surrounding nuclei. However, the computational cost of FCI calculations grows exponentially as the system size increases (the curse of dimensionality). Sophisticated methods have been developed to approximately solve the many-body Schr\"odinger equation and to provide insight into strongly-correlated electronic states, such as dynamical mean-field theory \cite{Georges1996,Kotliar2006}, quantum Monte-Carlo \cite{Ceperley1986,Wagner2016} and various multi-reference quantum chemistry methods \cite{Lischka2018}. These approaches are in general much more computationally demanding than DFT. 

Quantum embedding theories are promising approaches to study strongly-correlated states in materials because they enable the use of high-level theories for selected degrees of freedom of the system (defined by a chosen active space) while treating the rest of the system (environment) within mean-field theories. Various embedding schemes have been proposed based on different fundamental quantities \cite{Sun2016}, such as the electron density \cite{Huang2006,Huang2011,Goodpaster2014,Jacob2014,Genova2014,Wen2019}, density matrix \cite{Knizia2012,Wouters2016,pham2019periodic} and Green’s function \cite{Nguyen2016,Dvorak2019,Zhu2019,Aryasetiawan2004,Aryasetiawan2009,Miyake2009,Imada2010,Hirayama2013,Hirayama2017,Cho2018,Romanova2020}. In this work, we focus on a quantum embedding theory based on the screened Coulomb interactions, with the goal of developing an approach scalable to large systems, with hundreds of atoms. The method is inspired by the constrained random phase approximation (cRPA) \cite{Aryasetiawan2004}, which has been used to construct effective Hamiltonians acting on selected energy bands (active space) in complex materials, such as oxides \cite{Aryasetiawan2006,Shih2012,Nilsson2017,Tadano2019}, 2D materials \cite{Wehling2011}, and spin-defects in wide-gap semiconductors \cite{Bockstedte2018}. Two approximations are usually adopted to evaluate the dielectric screening in cRPA calculations: (i) the random phase approximation (RPA), which neglects the exchange-correlation interaction between electrons in evaluating the dielectric screening \cite{Honerkamp2018}, thus affecting the accuracy of the calculation; (ii) the Adler-Wiser formalism \cite{Adler1962,Wiser1963}
that involves explicit summations over empty states, thus affecting the efficiency of calculations. Recently, we developed a quantum embedding theory that overcomes both approximations \cite{Ma2020}: the dielectric screening is computed beyond the RPA by including exchange-correlation effects evaluated using a finite-field algorithm \cite{Ma2018,Nguyen2019}, and the summation over empty states is circumvented by using a compact basis obtained through the spectral decomposition of density response functions \cite{Wilson2008,Nguyen2012,Pham2013,Govoni2015}. We applied the embedding theory to study several spin-defects in semiconductors relevant for quantum information technologies \cite{Ma2020,Ma2020pccp}, and we demonstrated that the theory can accurately predict the excitation energies for strongly-correlated excited states.

In this work, we present a detailed derivation of the quantum embedding theory introduced in Ref. \citenum{Ma2020}, and we discuss the accuracy of several strategies to compute the dielectric screening of the environment beyond the random phase approximation (RPA). Using spin-defects as examples, we show that the most accurate results are obtained by properly including in the calculation the exchange-correlation effects of the environment without double counting exchange-correlation effects in the active space. We further extend the quantum embedding formalism to cases where the embedding is achieved by using projection operators that do not commute with the Kohn-Sham Hamiltonian. The generalized formulation is thus not restricted to the use of Kohn-Sham orbitals to define active spaces, but allows as well for the use of localized orbitals, e.g. maximally localized Wannier functions (MLWFs) \cite{marzari2012maximally}. We report proof-of-principles calculations of spin-defects and of a transition metal oxide, \ce{SrTiO3} using active spaces defined with MLWFs. 

The rest of the paper is organized as follows. In Sec. 2 we describe the formalism of the quantum embedding theory within and beyond the RPA description of dielectric screening, and we discuss how quantum embedding calculations can be performed without explicit summation over empty states. In Sec. 3 we present benchmark calculations on spin-defects and \ce{SrTiO3}. Sec. 4 contains our summary and conclusions.


\section{Methods}

In the following discussion, we will focus on systems represented by large periodic cells for which only the $\Gamma$ point is required to sample the Brillouin zone.

\subsection{Effective Hamiltonian and cRPA embedding}
Under the Born-Oppenheimer and nonrelativisitic approximations, the many-body Hamiltonian of a system of interacting electrons is
\begin{equation} \label{Hfull}
    H = \sum_{ij} t_{ij} a_{i}^{\dagger} a_{j} + \frac{1}{2} \sum_{ijkl} v_{ijkl} a_{i}^{\dagger} a_{j}^{\dagger} a_{l} a_{k}
\end{equation}
where $a^{\dagger}$ and $a$ are creation and annihilation operators acting on single-electron states $i, j, k, l$; the one-electron term  $t$ includes the kinetic energy and the electron-nuclei interaction; the two-electron term $v$ represents the \textit{bare} Coulomb interaction between electrons. The exact diagonalization of $H$ can only be carried out for small systems due to the high computational cost. 

If electronic excitations of interest occur within a small subspace (denoted as the active space A) of the full Hilbert space, then a quantum embedding scheme may be used to construct an effective Hamiltonian, $H^{\text{eff}}$, that acts only on the active space:
\begin{equation} \label{Heff}
    H^{\text{eff}} = \sum_{ij}^{\text{A}} t^{\text{eff}}_{ij} a_{i}^{\dagger} a_{j} + \frac{1}{2} \sum_{ijkl}^{\text{A}} v^{\text{eff}}_{ijkl} a_{i}^{\dagger} a_{j}^{\dagger} a_{l} a_{k}
\end{equation}
where $t^{\text{eff}}$ and $v^{\text{eff}}$ are renormalized one-electron and two-electron terms that take into account the interactions between the active space and the environment. The parameters of the effective Hamiltonian may be determined by fitting experimental results or may be derived from first-principles calculations. In this work we determine the parameters of the effective Hamiltonian using DFT calculations. We note again that the form of the effective Hamiltonian is general and may reduce to that of the Hubbard model if certain terms are excluded from the summations entering Eq.~\ref{Heff}, indicating that first-principles calculations of $v^{\text{eff}}_{ijkl}$ may be used to obtain Hubbard parameters (e.g., $U$) \cite{Aryasetiawan2004} for DFT+U calculations \cite{Anisimov1997,Timrov2018}, as discussed later in the paper.

Within DFT, a mean-field description of the full system is obtained by solving self-consistently the Kohn-Sham equations
\begin{equation} \label{KS}
    H_{\text{KS}} \psi_{m}(\bm{x}) = \varepsilon_{m} \psi_{m}(\bm{x}),
\end{equation}
where the Kohn-Sham Hamiltonian $H_{\text{KS}} = T + V_{\text{SCF}} = T + V_{\text{ion}} + V_{\text{H}} + V_{\text{xc}}$; $T$ is the kinetic energy operator; $V_{\text{SCF}}$ is the KS potential that includes the ionic $V_{\text{ion}}$, the Hartree $V_{\text{H}}$ and the exchange-correlation potential $V_{\text{xc}}$; $m$ is the index for Kohn-Sham (spin-)orbitals $\psi_{m}(\bm{x})$. $\bm{x} = (\bm{r}, \sigma)$ where $\bm{r}$ and $\sigma$ are electron coordinate and spin, respectively.
Within linear response, we define the Kohn-Sham polarizability $\chi_0(\bm{x}_1, \bm{x}_2, \omega)$ that represents the response of the electronic density at $\bm{x}_1$ caused by a monochromatic perturbative potential exerted at $\bm{x}_2$ with frequency $\omega$:
\begin{equation} \label{chi0AWrs}
\begin{split}
    \chi_0(\bm{x}_1, \bm{x}_2, \omega) = \sum_i^{\text{occ}} \sum_j^{\text{emp}} \psi_i(\bm{x}_1) \psi_j(\bm{x}_1) \psi_j(\bm{x}_2) \psi_i(\bm{x}_2) \left\{ \frac{1}{\omega - (\varepsilon_j - \varepsilon_i) + i\eta} - \frac{1}{\omega + (\varepsilon_j - \varepsilon_i) - i\eta} \right\}
\end{split}
\end{equation}
where ``occ" and ``emp" denote summation over occupied and empty (virtual) Kohn-Sham orbitals, respectively; $\eta$ is a infinitesimal positive value. Similar to the derivation of the \textit{GW} method within many-body perturbation theory \cite{Hedin1965}, one can use the polarizability to define the screened Coulomb interaction $W$ of the system. 
Within the RPA, $W$ can be computed through a Dyson equation:
\begin{equation} 
    W_{\text{rpa}} = v + v \chi_0 W_{\text{rpa}}
\end{equation}
where $v$ is the bare Coulomb interaction. 

In the cRPA approach, the two-body term of the effective Hamiltonian, $v^{\text{eff}}$, is computed as a partially screened Coulomb interaction
\begin{equation}
    W^\text{E}_{\text{rpa}} = v + v \chi_0^\text{E} W^\text{E}_{\text{rpa}}
\end{equation}
where E denotes the environment and $\chi_0^\text{E} = \chi_0 - \chi_0^\text{A}$. $\chi_0^\text{A}$ is the Kohn-Sham polarizability projected onto an active space \bibnote{Some literatures define $\chi_0^\text{A}$ by adding $O^\text{A}$ to all the four appearances of Kohn-Sham orbitals in Eq. \ref{chi0A}. This definition is equivalent to our definition if the active space is spanned by a set of Kohn-Sham orbitals. For general active spaces, we tested quantum embedding calculations using both definitions and we found the difference in results (e.g. Hubbard parameters of \ce{SrTiO3}) is negligible.}
\begin{equation} \label{chi0A}
    \chi_0^\text{A}(\bm{x}_1, \bm{x}_2, \omega) = \sum_{i}^{\text{occ}} \sum_{j}^{\text{emp}} (O^\text{A}\psi_i)(\bm{x}_1) (O^\text{A}\psi_j)(\bm{x}_1) \psi_j(\bm{x}_2) \psi_i(\bm{x}_2) \left\{ \frac{1}{\omega - (\varepsilon_j - \varepsilon_i) + i\eta} - \frac{1}{\omega + (\varepsilon_j - \varepsilon_i) - i\eta} \right\}
\end{equation}
where we defined the projector into the active space as $O^\text{A} = \sum_i^\text{A} \ket{\zeta_i} \bra{\zeta_i}$, and $\zeta_i$ are orthogonal orbitals spanning the active space.

If the active space is spanned by a set of Kohn-Sham orbitals, $\chi_0^\text{A}$ can be evaluated using a similar summation-over-state expression as Eq. \ref{chi0AWrs}, with summations limited to Kohn-Sham orbitals in the active space A only:
\begin{equation} \label{chi0Aks}
    \chi_0^\text{A}(\bm{x}_1, \bm{x}_2, \omega) = \sum_{i \in \text{A}}^{\text{occ}} \sum_{j \in \text{A}}^{\text{emp}} \psi_i(\bm{x}_1) \psi_j(\bm{x}_1) \psi_j(\bm{x}_2) \psi_i(\bm{x}_2) \left\{ \frac{1}{\omega - (\varepsilon_j - \varepsilon_i) + i\eta} - \frac{1}{\omega + (\varepsilon_j - \varepsilon_i) - i\eta} \right\}
\end{equation}

We note that $W^\text{E}_{\text{rpa}}$ represents the effective interaction between electrons considering only the screening of the environment, characterized by $\chi_0^\text{E}$; the full $W_{\text{rpa}}$ may be recovered by adding to $W^\text{E}_{\text{rpa}}$ the screening of the active space
\begin{equation} \label{wrpafull}
    W_{\text{rpa}} = W^\text{E}_{\text{rpa}} + W^\text{E}_{\text{rpa}} \chi_0^\text{A} W_{\text{rpa}}
\end{equation}

The two-body term $v^{\text{eff}}$ in the effective Hamiltonian (Eq. \ref{Heff}) may be computed as
\begin{equation}
    v^{\text{eff}}_{ijkl} = \int \mathrm{d}\bm{x} \mathrm{d}\bm{x}' \zeta_i(\bm{x}) \zeta_k(\bm{x}) W^\text{E}_{\text{rpa}}(\bm{x}, \bm{x}') \zeta_j(\bm{x}')\zeta_l(\bm{x}')
\end{equation}
As we discuss in Sec. \ref{beyondrpa}, $v^{\text{eff}}$ may be evaluated beyond the RPA, and calculations may be performed with general definitions of active spaces, without explicit summations over empty states (see Sec. \ref{noemptystates}).

Once $v^{\mathrm{eff}}$ is obtained, the one-body term $t^{\text{eff}}$ entering Eq.~\ref{Heff} may be computed by subtracting from the Kohn-Sham Hamiltonian a term that accounts for electrostatic and exchange-correlation interactions in the active space \cite{Bockstedte2018,Ma2020}
\begin{equation}
    t^{\mathrm{eff}}_{ij} = H^{\mathrm{KS}}_{ij} - \left( \sum_{kl} v^{\mathrm{eff}}_{ikjl} \rho_{kl} - \sum_{kl} v^{\mathrm{eff}}_{ijkl} \rho_{kl} \right)
\label{dc}
\end{equation}
where $\rho_{ij} = \sum_{m}^\text{occ} \braket{\zeta_i | \psi_m} \braket{\psi_m | \zeta_j}$ is the one-electron reduced density matrix. We remark that the strategy used in Eq. \ref{dc} to remove double counting is similar to the fully-localized limit (FLL) scheme \cite{Liechtenstein1995,Ryee2018} that is widely used in DFT+U calculations for treating double counting of interaction energies within $d$ or $f$ shells.

\subsection{Screened Coulomb interaction beyond the RPA}
\label{beyondrpa}

In this section we consider several definitions of partially screened Coulomb interaction $W^\text{E}$ beyond the RPA. Within many-body perturbation theory, two common expressions \bibnote{In principle, one can also define an electron-test-charge screened Coulomb interaction $W_{\text{etc}}$ that represents the screened interaction between a test charge and an electron. However, it is difficult to apply the cRPA-type treatment to $W_{\text{etc}}$ and define a partially screened interaction because it is difficult to write $W_{\text{etc}}$ in the form of a Dyson-like equation, so we will not consider $W_{\text{etc}}$ in this work.} of the screened Coulomb interaction beyond RPA are defined, depending on whether the interacting charges are excluded from or included in the self-consistent solution of the Kohn-Sham system. The two definitions are called test-charge $W$ ($W_{\text{tc}}$) and test-electron $W$ ($W_{\text{el}}$), respectively\cite{Hybertson1986,DelSole1994,Martin2016}:
\begin{equation} \label{WTC}
    W_{\text{tc}} = v + v P W_{\text{tc}}
\end{equation}
\begin{equation} \label{WE}
    W_{\text{el}} = f_{\text{Hxc}} + f_{\text{Hxc}} \chi_0 W_{\text{el}}
\end{equation}
where the irreducible polarizability $P$ is defined as the derivative of the electron density with respect to the classical electrostatic potential $V_{\text{cl}}$ (the sum of Hartree and external potential): $P = \frac{\delta n}{\delta V_{\text{cl}}} = \chi_0 + \chi_0 f_{\text{xc}} P$; the Hartree-exchange-correlation kernel $f_{\text{Hxc}}$ is defined as $ f_{\text{Hxc}} = v + f_{\text{xc}}$, where the exchange-correlation kernel $f_{\text{xc}}$ is the functional derivative of $V_{\text{xc}}$ with respect to the density $f_{\text{xc}} = \frac{\delta V_{\text{xc}}}{\delta n}$. $W_{\text{tc}}$ and $W_{\text{el}}$ are adopted in many-body perturbation theory calculations of quasiparticle energies and electron-phonon coupling beyond the RPA \cite{DelSole1994,Paier2008,Gruneis2014,Ma2018,McAvoy2018,Giustino2017}. $W_{\text{tc}}$ and $W_{\text{el}}$ both reduce to $W_{\text{rpa}}$ when the RPA is invoked, i.e. when $f_{\text{xc}} = 0$.

Inspired by the cRPA formalism, we define partially screened test-charge and test-electron Coulomb interactions ($W^\text{E}_{\text{tc}}$ and $W^\text{E}_{\text{el}}$, respectively). To this end, we partition the irreducible polarizability $P = P^\text{A} + P^\text{E}$ and the Kohn-Sham polarizability $\chi_0 = \chi_0^\text{A} + \chi_0^\text{E}$ into contributions from the active space and the environment, where $\chi_0^\text{A}$ is defined in Eq. \ref{chi0A}. We define $P^\text{A}$ as $P^\text{A} = \chi_0^\text{A} + \chi_0^\text{A} f_{\text{xc}} P^\text{A}$, and $W^\text{E}_{\text{tc}}$ and $W^\text{E}_{\text{el}}$ as:
\begin{equation}
    W^\text{E}_{\text{tc}} = v + v P^\text{E} W^\text{E}_{\text{tc}}
\end{equation}
\begin{equation}
    W^\text{E}_{\text{el}} = f_{\text{Hxc}} + f_{\text{Hxc}} \chi_0^\text{E} W^\text{E}_{\text{el}}
\label{weel}
\end{equation}

The quantities $W^\text{E}_{\text{tc}}$ and $W^\text{E}_{\text{el}}$ defined above have the property that the corresponding full $W$ can be obtained by further screening $W^\text{E}$ with $\chi_0^\text{A}$ or $P^\text{A}$, similar to the cRPA formulation (Eq. \ref{wrpafull}):
\begin{equation}
    W_{\text{tc}} = W^\text{E}_{\text{tc}} + W^\text{E}_{\text{tc}} P^\text{A} W_{\text{tc}}
\end{equation}
\begin{equation}
    W_{\text{el}} = W^\text{E}_{\text{el}} + W^\text{E}_{\text{el}} \chi_0^\text{A} W_{\text{el}}
\end{equation}

\begin{figure}[!h]
  \centering
  \includegraphics[width=6in]{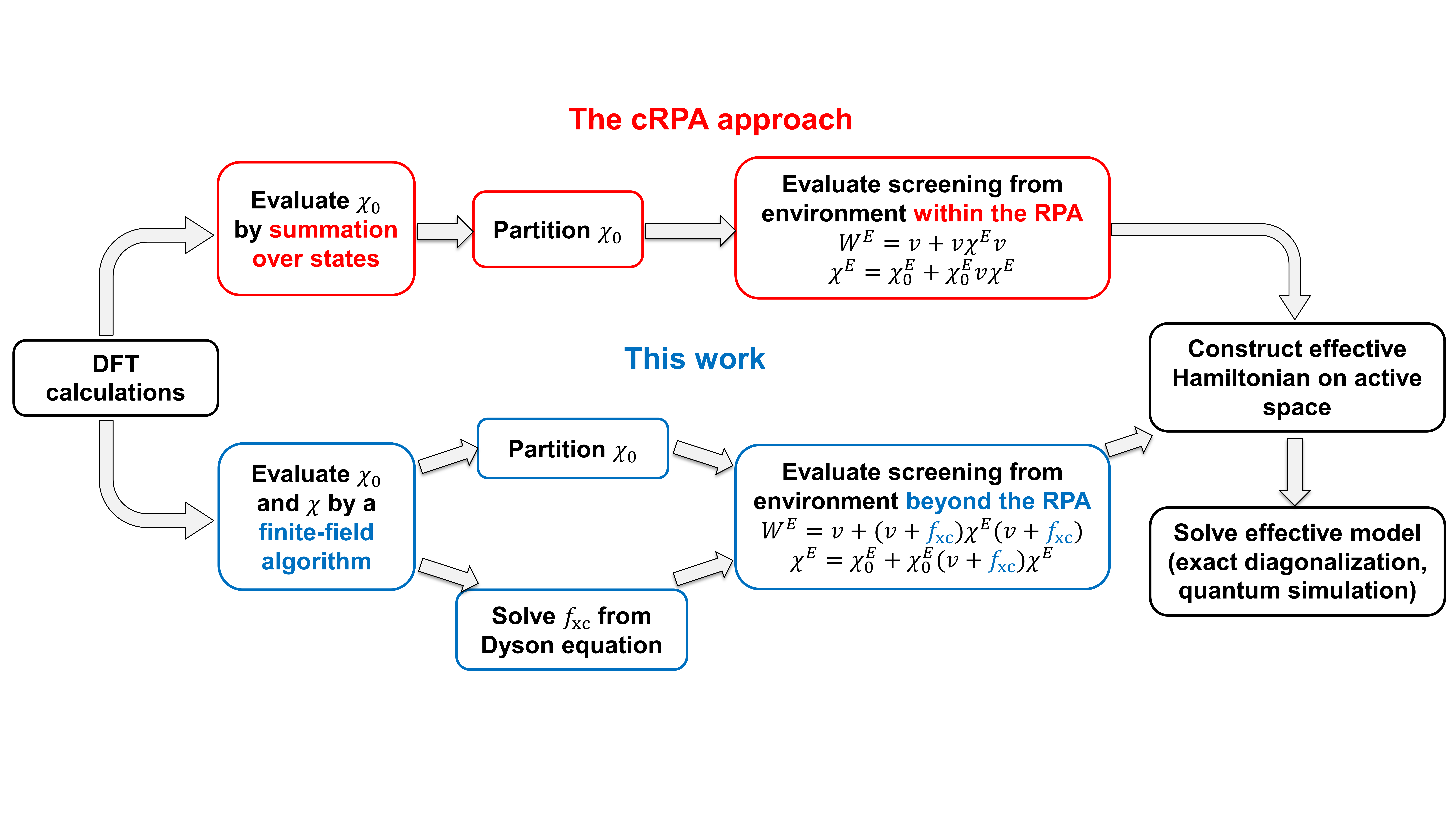}
  \caption{Comparison of workflows for quantum embedding calculations within and beyond the random phase approximation.}
  \label{workflow}
\end{figure}

We note that neither $W^\text{E}_{\text{tc}}$ nor $W^\text{E}_{\text{el}}$ shall be used as the effective electron interactions $v^{\text{eff}}$ beyond the RPA in the definition of the effective Hamiltonian. If $v^{\text{eff}} = W^\text{E}_{\text{tc}}$, then electrons in the active space interact with electrons in the environment through the bare Coulomb interaction, and the exchange-correlation effects between electrons in the active space and the environment are not correctly included. If $v^{\text{eff}} = W^\text{E}_{\text{el}}$, then the exchange-correlation effects are double-counted when diagonalizing the effective Hamiltonian, since the exchange-correlation effects between electrons in the active space are already accounted for through the $f_{\text{xc}}$ term entering the bare part of $W^\text{E}_{\text{el}}$. Hence, based on physical considerations, we propose to use $v^{\text{eff}}=W_{\text{vel}}$, where we define $W_{\text{vel}}$ as the sum of the bare Coulomb interaction $v$ and the  polarization part (second term in r.h.s. of Eq. \ref{weel}) of $W_{\text{el}}$:
\begin{equation}
    W^\text{E}_{\text{vel}} = v + f_{\text{Hxc}} \chi_0^\text{E} W^\text{E}_{\text{el}}
    \label{wmatrix}
\end{equation}
$W^\text{E}_{\text{vel}}$ includes exchange-correlation effects in the environment while avoiding the double counting of exchange-correlation effects in the active space (see also the Supporting Information (SI) of Ref. \citenum{Ma2020}). In Sec. \ref{w_benchmark} we perform quantum embedding calculations of spin-defects in semiconductors with effective interactions defined as $W^\text{E}_{\text{rpa}}$, $W^\text{E}_{\text{tc}}$, $W^\text{E}_{\text{el}}$ and $W^\text{E}_{\text{vel}}$, and we demonstrate that effective Hamiltonians constructed from $W^\text{E}_{\text{vel}}$ indeed lead to accurate predictions of excitation energies of spin-defects. To summarize the formalism described in this section, in Fig. \ref{workflow} we present a comparison of quantum embedding calculation workflows within and beyond the RPA using $W^\text{E}_{\text{rpa}}$ and $W^\text{E}_{\text{vel}}$, respectively.

\subsection{Evaluation of $W^\text{E}$ without empty states}
\label{noemptystates}

In this section we discuss how to evaluate $W^\text{E}$ without explicitly computing virtual electronic states, by representing $\chi_0$ on a compact basis. The formalism discussed in this subsection is general and is applicable regardless of whether the active space is defined using eigenstates of the Kohn-Sham Hamiltonian.

We consider the spin-unresolved $\chi_0 (\bm{r}, \bm{r}')$, where the spin index is already summed over. We use a so-called projected-dielectric eigendecomposition (PDEP) basis \cite{Wilson2008,Nguyen2012,Pham2013,Govoni2015}, obtained by the spectral decomposition of the symmetrized Kohn-Sham polarizability $\tilde{\chi}_0=v_{\text{c}}^{\frac{1}{2}} \chi_0 v_{\text{c}}^{\frac{1}{2}}$ at zero frequency. To obtain the PDEP basis, we use the Davidson algorithm to compute ${N_{\text{PDEP}}}$ eigenvectors of $\tilde{\chi}_0 (\omega = 0)$ with the most negative eigenvalues:
\begin{equation} \label{pdep}
    \tilde{\chi}_0 (\omega = 0) \approx \sum_i^{N_{\text{PDEP}}} \lambda_i \ket{\varphi_i} \bra{\varphi_i}
\end{equation}
The accuracy of response functions and screened Coulomb interactions are thus determined by the parameter ${N_{\text{PDEP}}}$ used in Eq. \ref{pdep}, and in the SI we show, as examples, the rapid convergence of excitation energies of spin-defects and Hubbard parameters of \ce{SrTiO3} as functions of ${N_{\text{PDEP}}}$.

We use a finite-field algorithm \cite{Ma2018, Nguyen2019} to evaluate the matrix elements of the symmetrized static reducible polarizability $\tilde{\chi} = \tilde{\chi}_0 + \tilde{\chi}_0 \tilde{f}_{\text{Hxc}} \tilde{\chi}$ on the PDEP basis, where $\tilde{f}_{\text{Hxc}} = v_{\text{c}}^{-\frac{1}{2}} f_{\text{Hxc}} v_{\text{c}}^{-\frac{1}{2}}$. In the finite-field algorithm, one evaluates the charge density $\rho^\pm_i$ of the Kohn-Sham system subject to the perturbation $\tilde{\varphi}_i=v_{\text{c}}^{\frac{1}{2}}\varphi_i$ by solving the Kohn-Sham equations with $H^{\mathrm{KS}} \pm \tilde{\varphi}_i$. Then one can compute the linear variation of the electronic density using central finite differences, i.e., $\Delta\rho_i=\frac{\rho^+_i-\rho^-_i}{2}$. The matrix elements of $\tilde{\chi}$ in the PDEP basis are therefore given by: 
\begin{equation}
    \tilde{\chi}_{ij} = \braket{\tilde{\varphi}_i|\Delta\rho_j}
\end{equation}
We note that all potential terms within the Kohn-Sham Hamiltonian are updated during the self-consistent iterations\cite{Ma2018, Nguyen2019}. We then compute the matrix elements of $\tilde{f}_{\text{xc}}$ using the following matrix identity:
\begin{equation} \label{fxcmatrix}
    \tilde{f}_{\text{xc}} = {\tilde{\chi}_{0}}^{-1} - \tilde{\chi}^{-1} - 1.
\end{equation}
The matrix inversion operations entering Eq.~\ref{fxcmatrix} is efficiently carried out using the PDEP compact basis set, where $N_{\text{PDEP}}\ll N_{\text{PW}}$ (the number of plane-wave basis functions).

At arbitrary frequency, the matrix elements of $\tilde{\chi}_0$ on the PDEP basis can be computed as\cite{Govoni2015}
\begin{equation} \label{chi0body}
    \tilde{\chi}_{0ij}(\omega) = \sum_n^{\text{occ}} \bra{\xi^i_{n}}  O^c \bigg [ (\varepsilon_{n}-H_{\text{KS}} - \omega + i\eta)^{-1} + (\varepsilon_{n}-H_{\text{KS}} + \omega + i\eta)^{-1}\bigg ] O^c \ket{\xi^j_{n}} \\
\end{equation}
where $i, j$ are indices for PDEP basis functions, and $\xi^i_{n}(\bm{r}) = \tilde{\varphi}_i(\bm{r}) \psi_{n}(\bm{r})$, where $\psi_{n}(\bm{r})$ denotes the spatial part of the Kohn-Sham orbital $\psi_{n}(\bm{x})$. The summation over empty states in Eq. \ref{chi0AWrs} is formally replaced by the projection operator $O^c$ onto virtual manifold. In practical calculations, $O^c$ is replaced with $1 - O^v$, where $O^v$ is the projection onto the occupied manifold. Thus Eq. \ref{chi0body} can be evaluated without explicitly summing over empty states. Furthermore, the frequency dependence of $\chi_0$ can be efficiently included through the use of the Lanczos algorithm \cite{Govoni2015}, although in this work we only consider zero frequency response functions and we do not further investigate their frequency dependence.

Similar to Eq. \ref{chi0body}, $\tilde{\chi}^\text{A}_0$ can be written as
\begin{equation} \label{chi0Abody}
    \tilde{\chi}^\text{A}_{0ij}(\omega) = \sum_n^{\text{occ}} \bra{\xi^{Ai}_{n}} O^\text{A} O^c \bigg [(\varepsilon_{n}-H_{\text{KS}} - \omega + i\eta)^{-1} + (\varepsilon_{n}-H_{\text{KS}} + \omega + i\eta)^{-1}\bigg ] O^c \ket{\xi^j_{n}}
\end{equation}
where $\xi^{Ai}_{n}(\bm{r}) = \tilde{\varphi}_i(\bm{r}) (O^\text{A}\psi_{n})(\bm{r})$. Therefore, $\tilde{\chi}^\text{A}_0$ at zero or finite frequency can be evaluated in a similar manner as $\tilde{\chi}_0$, i.e., without explicit summation over empty states. We note that Eq. \ref{chi0Abody} does not assume that the active space is defined through a set of Kohn-Sham orbitals (i.e. we do not assume that  $O^\text{A}$ commutes with $H^\text{KS}$).

Once $\tilde{\chi}^\text{A}_0$ is evaluated, we can compute a set of partial reducible polarizabilities
\begin{equation}
    \tilde{\chi}^\text{E}_{\text{rpa}} = \tilde{\chi}^\text{E}_0 + \tilde{\chi}^\text{E}_0 \tilde{\chi}^\text{E}_{\text{rpa}}
\end{equation}
\begin{equation}
    \tilde{\chi}^\text{E}_{\text{tc}} = \tilde{P}^\text{E} + \tilde{P}^\text{E} \tilde{\chi}^\text{E}_{\text{tc}}
\end{equation}
\begin{equation}
    \tilde{\chi}^\text{E}_{\text{el}} = \tilde{\chi}^\text{E}_0 + \tilde{\chi}^\text{E}_0 \tilde{f}_{\text{Hxc}} \tilde{\chi}^\text{E}_{\text{el}}
\end{equation}
that we use to define $W^\text{E}$ (note that $\chi = v_{\text{c}}^{-\frac{1}{2}} \tilde{\chi} v_{\text{c}}^{-\frac{1}{2}}$, where $\chi$ and $\tilde{\chi}$ are unsymmetrized and symmetrized response functions):
\begin{equation}
    W^\text{E}_{\text{rpa}} = v_{\text{c}} + v_{\text{c}} \chi^\text{E}_{\text{rpa}} v_{\text{c}}
\label{wrpa}
\end{equation}
\begin{equation}
    W^\text{E}_{\text{tc}} = v_{\text{c}} + v_{\text{c}} \chi^\text{E}_{\text{tc}} v_{\text{c}}
\label{wtc}
\end{equation}
\begin{equation}
    W^\text{E}_{\text{el}} = f_{\text{Hxc}} + f_{\text{Hxc}} \chi^\text{E}_{\text{el}} f_{\text{Hxc}}
\label{wel}
\end{equation}
\begin{equation}
    W^\text{E}_{\text{vel}} = v_{\text{c}} + f_{\text{Hxc}} \chi^\text{E}_{\text{el}} f_{\text{Hxc}}
\label{wvel}
\end{equation}

To construct effective Hamiltonians beyond the RPA, one simply evaluates matrix elements of $W^\text{E}$ defined in Eq. \ref{wtc} - \ref{wvel} on the active space, similar to Eq. \ref{wmatrix}. The evaluation of matrix elements of the bare Coulomb interaction is straightforward and can be performed using the standard Gygi-Baldereschi scheme \cite{Gygi1986}. The polarization component can be computed with a resolution-of-identity (RI) technique using the PDEP basis. We provide the detailed expression of the RI calculation in the SI, and we note again that the numerical accuracy of the matrix elements is controlled by the size of the PDEP basis $N_\text{PDEP}$.


\section{Results}

\subsection{Computational setup}

The calculations of effective Hamiltonians were carried out with the WEST code \cite{Govoni2015}. For calculations beyond the RPA, the exchange-correlation kernel $f_{\text{xc}}$ was evaluated using a finite-field algorithm \cite{Ma2018,Nguyen2019} by coupling the WEST code with the Qbox code \cite{Gygi2008} in client-server mode. We used $N_\text{PDEP} = 512$ for the evaluation of the density response functions (Eq. \ref{pdep}), and convergence tests are presented in the SI. Kohn-Sham DFT orbitals were obtained with the Quantum Espresso code \cite{Giannozzi2009} and MLWFs with the Wannier90 code \cite{mostofi2008wannier90}. FCI calculations using effective Hamiltonians were performed with the PySCF code \cite{Sun2017}.

In our calculations we used plane-wave basis sets with a kinetic energy cutoff of 50/70 Ry in the calculations of spin-defects/\ce{SrTiO3}. The electron-ion interactions were represented by norm-conserving pseudopotentials from the SG15 library \cite{Schlipf2015}. Spin-defects in diamond/4H-SiC were modeled with 216-atom/200-atom supercells. The \ce{SrTiO3} solid was modeled with a 135-atom supercell. All calculations were performed with $\Gamma$-point sampling of the Brillouin zone, except the calculation of MLWFs for \ce{SrTiO3}, which was performed starting from non-self-consistent DFT calculations using a $4\times4\times4$ k-point mesh.

\subsection{Calculation of environment dielectric screening beyond the RPA}
\label{w_benchmark}

\begin{figure}[!h]
  \centering
  \includegraphics[width=4in]{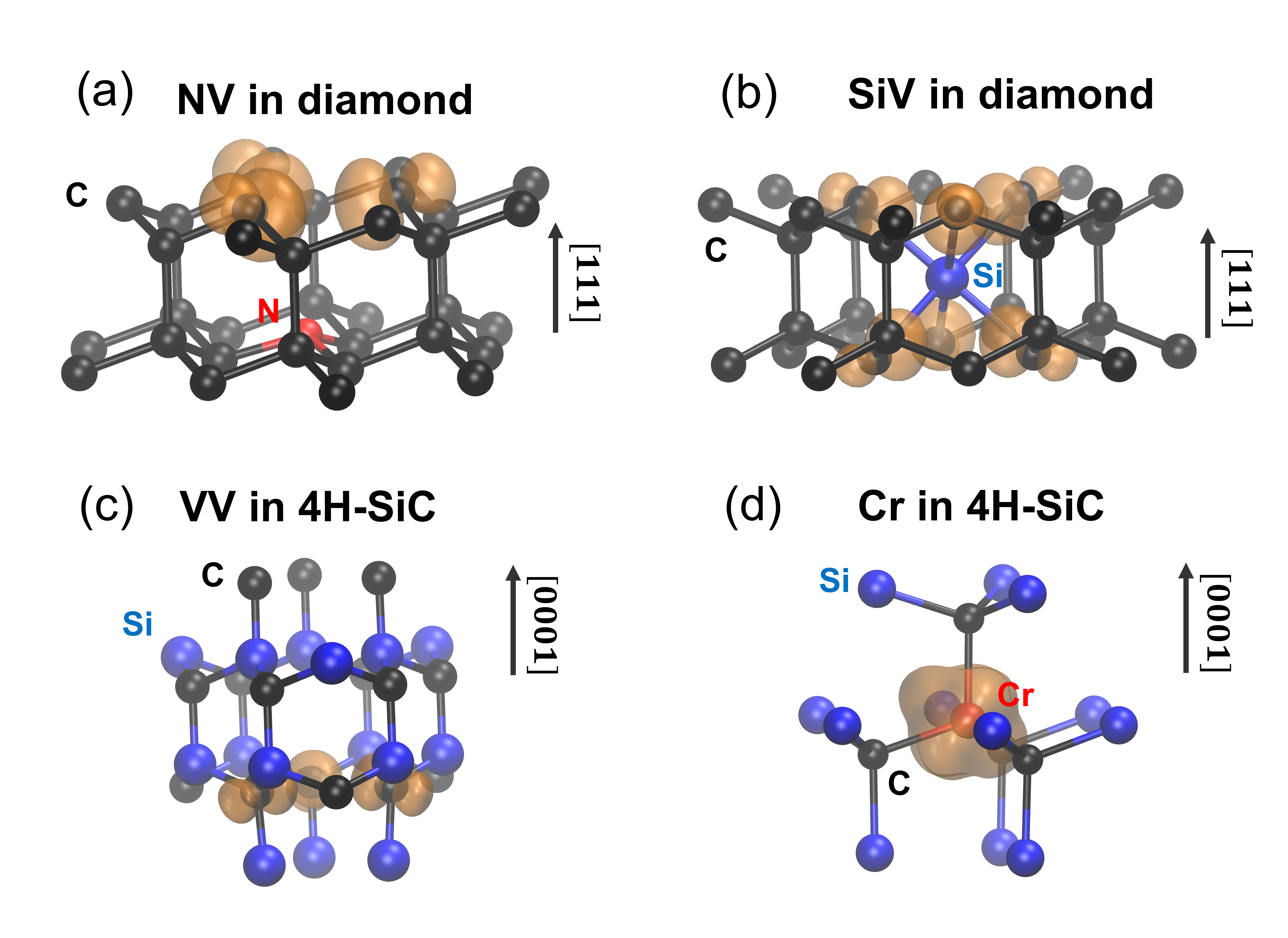}
  \caption{Structures and spin densities of the negatively-charged nitrogen-vacancy (NV) center in diamond (a), the neutral silicon-vacancy (SiV) center (b) in diamond, the neutral divacancy (VV) (\textit{hh} configuration) in 4H-SiC (c) and the chromium (4+) impurity (\textit{h} configuration) in 4H-SiC (d).}
  \label{defect_structures}
\end{figure}

We turn to discuss calculations of the environmental screened Coulomb interaction $W^\text{E}$ beyond the RPA, using the different expressions presented in Sec. \ref{beyondrpa} to compute the excitation energies of several spin-defects in wide-gap semiconductors: the negatively charged nitrogen-vacancy (NV) center and the neutral silicon-vacancy (SiV) center in diamond, and the neutral divacancy (VV) and Cr impurity in 4H-SiC. These spin-defects are promising platforms for realizing solid-state quantum bits for quantum information processing, and they possess spin-triplet ground state and exhibit strongly-correlated electronic states that are critical for the initialization and read-out of their spin states \cite{Weber2010,Seo2016,Seo2017,Ivady2018,Dreyer2018,Anderson2019}. For instance, it has been shown that the majority of low-lying many-body electronic states of the NV center are strongly-correlated \cite{Maze2011,Doherty2011}. We note that VV and Cr can exist in different configurations in the 4H-SiC lattice, and here we consider the hexagonal configurations. The atomistic structures of spin-defects are shown in Fig. \ref{defect_structures}.

\begin{figure}[!h]
  \centering
  \includegraphics[width=6.5in]{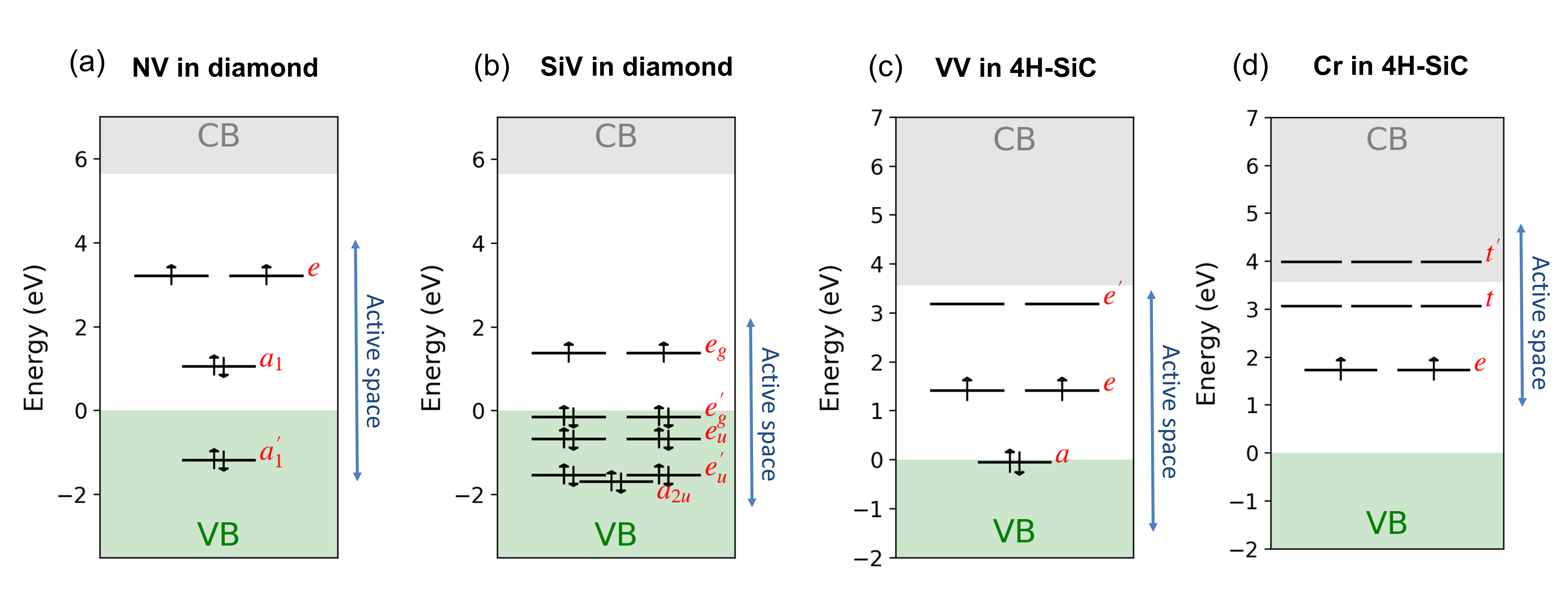}
  \caption{Defect levels obtained from spin-restricted DFT calculations using DDH functional for the negatively-charge nitrogen-vacancy (NV) center in diamond (a), the neutral silicon-vacancy (SiV) center (b) in diamond, the neutral divacancy (VV) in 4H-SiC (c) and the chromium (4+) impurity in 4H-SiC (d). Note that due to spin restriction, exchange splittings of defect levels in their spin triplet ground state are not reflected in these diagrams. The choice of active spaces for the NV and SiV centers is consistent with Ref. \citenum{Ma2020}. }
  \label{defect_levels}
\end{figure}

To compute the excitation energies of spin-defects using the quantum embedding theory, we first carried out ground state spin-unrestricted DFT calculations using the PBE functional \cite{Perdew1996} to optimize the structure of these defects. Then we performed spin-restricted DFT calculations using the PBE and a dielectric dependent hybrid (DDH) \cite{Skone2014,Skone2016,Brawand2016,Brawand2017,Gerosa_2017,Zheng2019} functional to obtain mean-field descriptions of the electronic states, providing starting points for the construction of effective Hamiltonians. The fraction of exact exchange used in DDH calculations is taken as the inverse of the high-frequency dielectric constants $\epsilon_\infty$,  which is self-consistently determined to be 5.61/6.57 for diamond/silicon carbide; for comparison the experimental values for $\epsilon_\infty$ 5.70/6.52\cite{Cardona2005}. The spin restriction ensures that both spin channels are treated on an equal footing and the eigenstates of the resulting effective Hamiltonians are eigenstates of $S^2$ \cite{Bockstedte2018,Ivady2020}. In Fig. \ref{defect_levels} we show the position of single-particle defect levels of spin-defects and the active spaces used to define the effective Hamiltonians. The active spaces are chosen to be sufficiently large to yield converged excitation energies of the spin-defects (see convergence tests reported in Ref. \citenum{Ma2020,Ma2020pccp}). Using the effective Hamiltonians constructed with the quantum embedding theory, we performed FCI calculations \cite{Knowles1984} to compute the low-energy eigenstates and vertical excitation energies of spin-defects.

In Table \ref{excitation_energies} we present vertical excitation energies obtained from FCI calculations on effective Hamiltonians constructed using different expression of $W^\text{E}$ ($W^\text{E}_{\text{rpa}}$, $W^\text{E}_{\text{tc}}$, $W^\text{E}_{\text{el}}$, $W^\text{E}_{\text{vel}}$) and with different DFT starting point (PBE, DDH). The $W^\text{E}_{\text{rpa}}$ and $W^\text{E}_{\text{vel}}$ results for NV and SiV were also reported in Ref. \citenum{Ma2020} and Ref. \citenum{Ma2020pccp}. As expected, excitation energies obtained using DDH starting points are significantly larger than those obtained using PBE, and DDH results are in better agreement with available reference values \cite{Ma2020pccp}. This comparison of DDH and PBE results is consistent with previous reports of $\Delta$-SCF calculations of triplet excited states of NV and VV \cite{Seo2017}.

Using calculations starting from Kohn-Sham eigenvalues and orbitals obtained with DDH functional, we obtain different results using different definitions of $W^\text{E}$. Compared to the results of cRPA calculations, calculations using $W^\text{E}_{\text{tc}}$ and $W^\text{E}_{\text{el}}$ lead to lower excitation energies while calculations using $W^\text{E}_{\text{vel}}$ lead to higher excitation energies. These differences can be understood by noticing that the exchange-correlation kernel $f_\text{xc}$ tends to cancel out a small portion of the bare Coulomb interaction $v$. By inspecting Eq. \ref{wrpa}-\ref{wvel}, one can see that compared to the cRPA case, the cancellation of the bare Coulomb interaction and the exchange-correlation kernel leads to a smaller effective interaction $W^\text{E}$ in the case of $W^\text{E}_{\text{tc}}$ and $W^\text{E}_{\text{el}}$ and a larger effective interaction in the case of $W^\text{E}_{\text{vel}}$.

By comparing the calculated excitation energies obtained with the DDH functional with experimental values, we find that using $W^\text{E}_{\text{vel}}$ leads to best overall agreement. As an example, we discuss in detail the excitation energies between ${}^1A_{1}$ and ${}^1E_{}$ states of the NV center. Based on molecular orbital theory analysis \cite{Maze2011,Doherty2011}, the ${}^1E_{} \rightarrow {}^1A_{1}$ excitation is a spin-flip transition of electrons in the $e$ defect orbitals, and its excitation energy is approximately equal to twice the exchange integral $\braket{e_xe_x|W^\text{E}|e_ye_y}$, where $e_x$ and $e_y$ are the two degenerate $e$ orbitals. Comparing the embedding theory results for excitation energies with the experimental zero phonon line (ZPL) value of 1.190 eV (which is a lower bound to the vertical excitation energy), we observe that embedding calculations using $W^\text{E}_{\text{rpa}}$ (cRPA calculations) leads to an underestimation of the excitation energy (0.900 eV). Going beyond the RPA by using $W^\text{E}_{\text{tc}}$ and $W^\text{E}_{\text{el}}$ leads to even more severe underestimations (0.471 eV and 0.824 eV). Instead, by properly accounting for the exchange-correlation effects between active space and environment, the excitation energy obtained with $W^\text{E}_{\text{vel}}$ (1.198 eV) significantly reduces the underestimation of the cRPA result and leads to the best agreement with experiment. Similar trends are observed in the comparison of other calculated excitation energies with experiment, as summarized in Table \ref{excitation_energies}.

\begin{table*}[!h]
\caption{Vertical excitation energies (eV) of the negatively charged nitrogen vacancy (NV) center and neutral silicon vacancy (SiV) center in diamond, and the neutral divacancy (VV) and Cr impurity (4+) in 4H-SiC. DFT calculations were performed using the PBE and the DDH functional. Quantum embedding calculations were performed with different definitions of $W^\text{E}$ (see text). Experimental measurements of zero phonon line (ZPL) energies are shown in brackets in the last column. Reference vertical excitation energies are computed from experimental ZPL when Stokes energies are available. }
\begin{tabular}{lllllllllll}
\hline
        &                                      & \multicolumn{4}{l}{PBE} & \multicolumn{4}{l}{DDH} &            Previous works \\
        &                                      & $W^\textrm{E}_\textrm{rpa}$ & $W^\textrm{E}_\textrm{tc}$ & $W^\textrm{E}_\textrm{el}$ & $W^\textrm{E}_\textrm{vel}$ & $W^\textrm{E}_\textrm{rpa}$ & $W^\textrm{E}_\textrm{tc}$ & $W^\textrm{E}_\textrm{el}$ & \multicolumn{2}{l}{$W^\textrm{E}_\textrm{vel}$} \\
System & Excitation &                             &                            &                            &                             &                             &                            &                            &                             &                \\
\hline
NV & ${}^3E_{} \leftrightarrow {}^3A_{2}$ &                       1.395$^a$ &                      1.282 &                      1.377 &                       1.458$^a$ &                       1.921$^b$ &                      1.775 &                      1.890 &                       2.001$^b$ &  2.180$^c$ (1.945$^c$) \\
       & ${}^1A_{1} \leftrightarrow {}^3A_{2}$ &                       1.211$^a$ &                      0.832 &                      1.162 &                       1.437$^a$ &                       1.376$^b$ &                      0.788 &                      1.274 &                       1.759$^b$ &                \\
       & ${}^1E_{} \leftrightarrow {}^3A_{2}$ &                       0.396$^a$ &                      0.305 &                      0.384 &                       0.444$^a$ &                       0.476$^b$ &                      0.317 &                      0.450 &                       0.561$^b$ &                \\
       & ${}^1A_{1} \leftrightarrow {}^1E_{}$ &                       0.815$^a$ &                      0.527 &                      0.778 &                       0.993$^a$ &                       0.900$^b$ &                      0.471 &                      0.824 &                       1.198$^b$ &        (1.190$^d$) \\
       & ${}^3E_{} \leftrightarrow {}^1A_{1}$ &                       0.184$^a$ &                      0.449 &                      0.215 &                       0.020$^a$ &                       0.545$^b$ &                      0.987 &                      0.616 &                       0.243$^b$ &  (0.344-0.430$^e$) \\
SiV & ${}^3E_{u} \leftrightarrow {}^3A_{2g}$ &                       1.247$^a$ &                      1.263 &                      1.244 &                       1.258$^a$ &                       1.590$^b$ &                      1.623 &                      1.586 &                       1.594$^b$ &   1.568$^f$ (1.31$^g$) \\
       & ${}^1E_{g} \leftrightarrow {}^3A_{2g}$ &                       0.232$^a$ &                      0.202 &                      0.223 &                       0.281$^a$ &                       0.261$^b$ &                      0.215 &                      0.244 &                       0.336$^b$ &                \\
       & ${}^1A_{1g} \leftrightarrow {}^3A_{2g}$ &                       0.404$^a$ &                      0.358 &                      0.391 &                       0.478$^a$ &                       0.466$^b$ &                      0.393 &                      0.440 &                       0.583$^b$ &                \\
       & ${}^1A_{1u} \leftrightarrow {}^3A_{2g}$ &                       1.262$^a$ &                      1.265 &                      1.258 &                         1.277$^a$ &                       1.608$^b$ &                      1.617 &                      1.602 &                       1.623$^b$ &                \\
VV & ${}^3E_{} \leftrightarrow {}^3A_{2}$ &                       0.962 &                      0.881 &                      0.957 &                       0.985 &                       1.231 &                      1.109 &                      1.220 &                       1.270 &       1.240$^h$ (1.094$^i$) \\
       & ${}^1A_{1} \leftrightarrow {}^3A_{2}$ &                       0.776 &                      0.483 &                      0.756 &                       0.908 &                       0.921 &                      0.429 &                      0.875 &                       1.156 &                \\
       & ${}^1E_{} \leftrightarrow {}^3A_{2}$ &                       0.246 &                      0.176 &                      0.242 &                       0.274 &                       0.299 &                      0.169 &                      0.288 &                       0.348 &                \\
       & ${}^1A_{1} \leftrightarrow {}^1E_{}$ &                       0.530 &                      0.308 &                      0.515 &                       0.634 &                       0.622 &                      0.260 &                      0.587 &                       0.808 &                \\
       & ${}^3E_{} \leftrightarrow {}^1A_{1}$ &                       0.186 &                      0.398 &                      0.201 &                       0.077 &                       0.310 &                      0.680 &                      0.345 &                       0.114 &                \\
Cr & ${}^3E_{} \leftrightarrow {}^3A_{2}$ &                       0.931 &                      0.984 &                      0.931 &                       0.890 &                       1.468 &                      1.587 &                      1.534 &                       1.343 &                \\
       & ${}^3A_{1} \leftrightarrow {}^3A_{2}$ &                       1.002 &                      1.072 &                      1.011 &                       0.956 &                       1.527 &                      1.664 &                      1.602 &                       1.394 &                \\
       & ${}^3E_{}' \leftrightarrow {}^3A_{2}$ &                       1.134 &                      1.122 &                      1.073 &                       1.189 &                       1.769 &                      1.760 &                      1.705 &                       1.744 &                \\
       & ${}^3A_{2}' \leftrightarrow {}^3A_{2}$ &                       1.177 &                      1.153 &                      1.106 &                       1.237 &                       1.805 &                      1.794 &                      1.731 &                       1.783 &                \\
       & ${}^1E_{} \leftrightarrow {}^3A_{2}$ &                       0.902 &                      0.854 &                      0.828 &                       1.017 &                       1.097 &                      1.075 &                      1.073 &                       1.130 &        (1.190$^j$) \\
       & ${}^1A_{1} \leftrightarrow {}^3A_{2}$ &                       1.467 &                      1.366 &                      1.374 &                       1.566 &                       1.992 &                      1.873 &                      1.819 &                       2.013 &                \\
\hline
\end{tabular}
\label{excitation_energies}
$^a$Ref \citenum{Ma2020pccp}.
$^b$Ref \citenum{Ma2020}.
$^c$Ref \citenum{Davies1976}.
$^d$Ref \citenum{Rogers2008}.
$^e$Estimated by Ref \citenum{Goldman2015} using a model for intersystem crossing.
$^f$Computed using Stokes energy from Ref \citenum{Thiering2019}.
$^g$Ref \citenum{Green2019}.
$^h$Computed using Stokes energy from Ref \citenum{Bockstedte2018}. 
$^i$Ref \citenum{Koehl2011}.
$^j$Ref \citenum{Son1999}.
\end{table*}

\subsection{Calculation of environment dielectric screening with general active spaces}

We now present two proof-of-principles examples of quantum embedding calculations using general active spaces.

In the first example, we compute the vertical excitation energies of the NV center in diamond using a minimum active space composed of 4 Kohn-Sham orbitals $a'_1$, $a_1$, $e_x$, and $e_y$. We performed quantum embedding calculations both by directly using the 4 Kohn-Sham orbitals and by using 4 MLWFs computed from the Kohn-Sham orbitals. The resulting excitation energies for the many-body ${}^1E_{}/{}^1A_{1}/{}^3E_{}$ states are 0.485/1.364/1.977 eV at the cRPA level if evaluated using Kohn-Sham orbitals and 0.485/1.364/1.976 eV if evaluated using MLWFs (beyond-RPA results show similar agreement). The excellent agreement between results obtained using the two different active spaces serves as a validation of the formalism and implementation.


In the second example, we apply the quantum embedding theory to compute the Hubbard U and J parameters entering DFT+U calculations \cite{Anisimov1997}. Hubbard parameters can be computed as the average values of certain matrix elements in the effective Hamiltonian, and the cRPA approach has been extensively used for first-principles predictions of these parameters for solids containing transition metal ions \cite{Solovyev2005,Aryasetiawan2006,Imada2010,Karlsson2010,Shih2012,Nilsson2017,Tadano2019}. 


Here we apply the quantum embedding theory to compute the Hubbard-U and Hubbard-J parameters for the Ti $t_{2g}$ orbitals of \ce{SrTiO3}. We first perform ground state DFT calculations of \ce{SrTiO3} using the PBE functional to obtain the eigenstates of the Kohn-Sham Hamiltonian. Then we construct effective Hamiltonians where the active space is composed of the three MLWFs that correspond to the $t_{2g}$ orbitals of Ti (see SI). This choice of active space corresponds to the $t_{2g} - t_{2g}$ model of \ce{SrTiO3} often considered in literature\cite{vaugier2012hubbard}. The Hubbard parameters can then be computed as $U = \frac{1}{3}\sum_{i=1}^3 W^{\textrm{E}}_{iiii}$ and $J = \frac{1}{6} \sum_{i,j=1,i\neq j}^3 W^{\textrm{E}}_{ijji}$. 




Quantum embedding calculations at the RPA level predict the Hubbard U/J parameters to be 3.90/0.58 eV, in good agreement with previous cRPA calculations (3.76/0.46 eV for U/J) \cite{hirayama2012ab}. The small differences are attributed to different choices of the pseudopotentials and Brilliouin zone sampling. The Hubbard-J parameters are found to be rather insensitive to the choice of $W^\text{E}$ (0.58/0.58/0.58/0.59 eV for $W^\text{E}_{\text{rpa}}$/$W^\text{E}_{\text{tc}}$/$W^\text{E}_{\text{el}}$/$W^\text{E}_{\text{vel}}$), as opposed to the Hubbard-U parameters (3.90/3.58/3.87/4.37 eV for $W^\text{E}_{\text{rpa}}$/$W^\text{E}_{\text{tc}}$/$W^\text{E}_{\text{el}}$/$W^\text{E}_{\text{vel}}$). We note that calculations using $W^\text{E}_{\text{tc}}$ and $W^\text{E}_{\text{el}}$ yield smaller values of U (3.58 eV and 3.87 eV), while calculations using $W^\text{E}_{\text{vel}}$ a larger value (4.37 eV), consistent with the trend observed for the excitation energies of spin-defects as discussed in Sec. \ref{w_benchmark}.

In order to assess the impact of the Hubbard parameters predicted by the embedding theory on the electronic structure of \ce{SrTiO3}, we performed DFT+U calculations to obtain the band gap of \ce{SrTiO3}. The experimental (indirect) band gap is 3.25 eV and the PBE prediction is 1.93 eV. The DFT+U calculations using Hubbard-U parameters with $W^\text{E}_{\text{rpa}}$/$W^\text{E}_{\text{tc}}$/$W^\text{E}_{\text{el}}$/$W^\text{E}_{\text{vel}}$ lead to band gaps of 2.25/2.22/2.25/2.30 eV, i.e. to values rather similar to each other.Although DFT+U calculations do not yield band gaps in quantitative agreement with experimental band gap due to covalent effects \cite{Ricca2020}, using $W^\text{E}_{\text{vel}}$ as the effective electron-electron interaction leads to slight improvement in the predicted gap value.

\section{Conclusions}

In summary, in this work we presented a detailed derivation of the quantum embedding theory recently introduced in Ref. \citenum{Ma2020}, and we generalized the formulation to active spaces defined using orbitals that are not eigenstates of the Kohn-Sham Hamiltonian. In addition, we discussed different approaches to compute the dielectric screening beyond the random phase approximation, which we applied to spin defects in semiconductors. In particular, we presented a physically motivated choice of the definition of screened Coulomb interaction, which turns out to yield the best agreement with experiments for excitation energies. Finally, using NV center in diamond and the \ce{SrTiO3} solid as examples, we demonstrate how quantum embedding calculations may be performed using general active spaces composed of localized orbitals.

We note that one of the possible advantages of using MLWFs instead of eigenstates is the ability to define active spaces associated to specific regions of the material, which can then be treated at different levels of theory. For example consider a nanocomposite (e.g. a nanoparticle in a matrix or in a solvent) or solvated ions or molecules in water. We may associate MLWFs to the nanoparticle or to the solute and its solvation shell, in order to define an active space and consider electronic excitations within that active space. The embedding theory is then used to properly take into account matrix or solvation effects.

Our work paves the way to the application of the quantum embedding theory to challenging chemical and materials science problems, which may be solved using effective Hamiltonians on classical and near-term quantum computers.

\section*{Acknowledgements}

This work was supported by MICCoM, as part of the Computational Materials Sciences Program funded by the U.S. Department of Energy, Office of Science, Basic Energy Sciences, Materials Sciences and Engineering Division through Argonne National Laboratory, under contract number DE-AC02-06CH11357. This research used resources of the National Energy Research Scientific Computing Center (NERSC), a DOE Office of Science User Facility supported by the Office of Science of the US Department of Energy under Contract No. DE-AC02-05CH11231, and resources of the University of Chicago Research Computing Center.

\pagebreak

\bibliography{ref}

\providecommand{\latin}[1]{#1}
\makeatletter
\providecommand{\doi}
  {\begingroup\let\do\@makeother\dospecials
  \catcode`\{=1 \catcode`\}=2 \doi@aux}
\providecommand{\doi@aux}[1]{\endgroup\texttt{#1}}
\makeatother
\providecommand*\mcitethebibliography{\thebibliography}
\csname @ifundefined\endcsname{endmcitethebibliography}
  {\let\endmcitethebibliography\endthebibliography}{}
\begin{mcitethebibliography}{99}
\providecommand*\natexlab[1]{#1}
\providecommand*\mciteSetBstSublistMode[1]{}
\providecommand*\mciteSetBstMaxWidthForm[2]{}
\providecommand*\mciteBstWouldAddEndPuncttrue
  {\def\EndOfBibitem{\unskip.}}
\providecommand*\mciteBstWouldAddEndPunctfalse
  {\let\EndOfBibitem\relax}
\providecommand*\mciteSetBstMidEndSepPunct[3]{}
\providecommand*\mciteSetBstSublistLabelBeginEnd[3]{}
\providecommand*\EndOfBibitem{}
\mciteSetBstSublistMode{f}
\mciteSetBstMaxWidthForm{subitem}{(\alph{mcitesubitemcount})}
\mciteSetBstSublistLabelBeginEnd
  {\mcitemaxwidthsubitemform\space}
  {\relax}
  {\relax}

\bibitem[Cohen \latin{et~al.}(2008)Cohen, Mori-S{\'a}nchez, and
  Yang]{Cohen2008}
Cohen,~A.~J.; Mori-S{\'a}nchez,~P.; Yang,~W. Insights into current limitations
  of density functional theory. \emph{Science} \textbf{2008}, \emph{321},
  792--794\relax
\mciteBstWouldAddEndPuncttrue
\mciteSetBstMidEndSepPunct{\mcitedefaultmidpunct}
{\mcitedefaultendpunct}{\mcitedefaultseppunct}\relax
\EndOfBibitem
\bibitem[Su \latin{et~al.}(2018)Su, Li, and Yang]{Su2018}
Su,~N.~Q.; Li,~C.; Yang,~W. Describing strong correlation with fractional-spin
  correction in density functional theory. \emph{Proc. Natl. Acad. Sci. U. S.
  A.} \textbf{2018}, \emph{115}, 9678--9683\relax
\mciteBstWouldAddEndPuncttrue
\mciteSetBstMidEndSepPunct{\mcitedefaultmidpunct}
{\mcitedefaultendpunct}{\mcitedefaultseppunct}\relax
\EndOfBibitem
\bibitem[Anisimov \latin{et~al.}(1997)Anisimov, Aryasetiawan, and
  Lichtenstein]{Anisimov1997}
Anisimov,~V.~I.; Aryasetiawan,~F.; Lichtenstein,~A. First-principles
  calculations of the electronic structure and spectra of strongly correlated
  systems: the LDA+ U method. \emph{J. Phys.: Condens. Matter} \textbf{1997},
  \emph{9}, 767\relax
\mciteBstWouldAddEndPuncttrue
\mciteSetBstMidEndSepPunct{\mcitedefaultmidpunct}
{\mcitedefaultendpunct}{\mcitedefaultseppunct}\relax
\EndOfBibitem
\bibitem[Bockstedte \latin{et~al.}(2018)Bockstedte, Sch\"utz, Garratt,
  Iv{\'{a}}dy, and Gali]{Bockstedte2018}
Bockstedte,~M.; Sch\"utz,~F.; Garratt,~T.; Iv{\'{a}}dy,~V.; Gali,~A. Ab initio
  description of highly correlated states in defects for realizing quantum
  bits. \emph{npj Quantum Mater.} \textbf{2018}, \emph{3}, 31\relax
\mciteBstWouldAddEndPuncttrue
\mciteSetBstMidEndSepPunct{\mcitedefaultmidpunct}
{\mcitedefaultendpunct}{\mcitedefaultseppunct}\relax
\EndOfBibitem
\bibitem[Kurashige \latin{et~al.}(2013)Kurashige, Chan, and
  Yanai]{Kurashige2013}
Kurashige,~Y.; Chan,~G. K.-L.; Yanai,~T. Entangled quantum electronic
  wavefunctions of the Mn4CaO5 cluster in photosystem {II}. \emph{Nat. Chem.}
  \textbf{2013}, \emph{5}, 660--666\relax
\mciteBstWouldAddEndPuncttrue
\mciteSetBstMidEndSepPunct{\mcitedefaultmidpunct}
{\mcitedefaultendpunct}{\mcitedefaultseppunct}\relax
\EndOfBibitem
\bibitem[Sharma \latin{et~al.}(2014)Sharma, Sivalingam, Neese, and
  Chan]{Sharma2014}
Sharma,~S.; Sivalingam,~K.; Neese,~F.; Chan,~G. K.-L. Low-energy spectrum of
  iron{\textendash}sulfur clusters directly from many-particle quantum
  mechanics. \emph{Nat. Chem.} \textbf{2014}, \emph{6}, 927--933\relax
\mciteBstWouldAddEndPuncttrue
\mciteSetBstMidEndSepPunct{\mcitedefaultmidpunct}
{\mcitedefaultendpunct}{\mcitedefaultseppunct}\relax
\EndOfBibitem
\bibitem[Georges \latin{et~al.}(1996)Georges, Kotliar, Krauth, and
  Rozenberg]{Georges1996}
Georges,~A.; Kotliar,~G.; Krauth,~W.; Rozenberg,~M.~J. Dynamical mean-field
  theory of strongly correlated fermion systems and the limit of infinite
  dimensions. \emph{Rev. Mod. Phys.} \textbf{1996}, \emph{68}, 13--125\relax
\mciteBstWouldAddEndPuncttrue
\mciteSetBstMidEndSepPunct{\mcitedefaultmidpunct}
{\mcitedefaultendpunct}{\mcitedefaultseppunct}\relax
\EndOfBibitem
\bibitem[Kotliar \latin{et~al.}(2006)Kotliar, Savrasov, Haule, Oudovenko,
  Parcollet, and Marianetti]{Kotliar2006}
Kotliar,~G.; Savrasov,~S.~Y.; Haule,~K.; Oudovenko,~V.~S.; Parcollet,~O.;
  Marianetti,~C.~A. Electronic structure calculations with dynamical mean-field
  theory. \emph{Rev. Mod. Phys.} \textbf{2006}, \emph{78}, 865--951\relax
\mciteBstWouldAddEndPuncttrue
\mciteSetBstMidEndSepPunct{\mcitedefaultmidpunct}
{\mcitedefaultendpunct}{\mcitedefaultseppunct}\relax
\EndOfBibitem
\bibitem[Ceperley and Alder(1986)Ceperley, and Alder]{Ceperley1986}
Ceperley,~D.; Alder,~B. Quantum Monte Carlo. \emph{Science} \textbf{1986},
  \emph{231}, 555--560\relax
\mciteBstWouldAddEndPuncttrue
\mciteSetBstMidEndSepPunct{\mcitedefaultmidpunct}
{\mcitedefaultendpunct}{\mcitedefaultseppunct}\relax
\EndOfBibitem
\bibitem[Wagner and Ceperley(2016)Wagner, and Ceperley]{Wagner2016}
Wagner,~L.~K.; Ceperley,~D.~M. Discovering correlated fermions using quantum
  Monte Carlo. \emph{Rep. Prog. Phys.} \textbf{2016}, \emph{79}, 094501\relax
\mciteBstWouldAddEndPuncttrue
\mciteSetBstMidEndSepPunct{\mcitedefaultmidpunct}
{\mcitedefaultendpunct}{\mcitedefaultseppunct}\relax
\EndOfBibitem
\bibitem[Lischka \latin{et~al.}(2018)Lischka, Nachtigallov{\'{a}}, Aquino,
  Szalay, Plasser, Machado, and Barbatti]{Lischka2018}
Lischka,~H.; Nachtigallov{\'{a}},~D.; Aquino,~A. J.~A.; Szalay,~P.~G.;
  Plasser,~F.; Machado,~F. B.~C.; Barbatti,~M. Multireference Approaches for
  Excited States of Molecules. \emph{Chem. Rev.} \textbf{2018}, \emph{118},
  7293--7361\relax
\mciteBstWouldAddEndPuncttrue
\mciteSetBstMidEndSepPunct{\mcitedefaultmidpunct}
{\mcitedefaultendpunct}{\mcitedefaultseppunct}\relax
\EndOfBibitem
\bibitem[Sun and Chan(2016)Sun, and Chan]{Sun2016}
Sun,~Q.; Chan,~G. K.-L. Quantum Embedding Theories. \emph{Acc. Chem. Res.}
  \textbf{2016}, \emph{49}, 2705--2712\relax
\mciteBstWouldAddEndPuncttrue
\mciteSetBstMidEndSepPunct{\mcitedefaultmidpunct}
{\mcitedefaultendpunct}{\mcitedefaultseppunct}\relax
\EndOfBibitem
\bibitem[Huang and Carter(2006)Huang, and Carter]{Huang2006}
Huang,~P.; Carter,~E.~A. Self-consistent embedding theory for locally
  correlated configuration interaction wave functions in condensed matter.
  \emph{J. Chem. Phys.} \textbf{2006}, \emph{125}, 084102\relax
\mciteBstWouldAddEndPuncttrue
\mciteSetBstMidEndSepPunct{\mcitedefaultmidpunct}
{\mcitedefaultendpunct}{\mcitedefaultseppunct}\relax
\EndOfBibitem
\bibitem[Huang \latin{et~al.}(2011)Huang, Pavone, and Carter]{Huang2011}
Huang,~C.; Pavone,~M.; Carter,~E.~A. Quantum mechanical embedding theory based
  on a unique embedding potential. \emph{J. Chem. Phys.} \textbf{2011},
  \emph{134}, 154110\relax
\mciteBstWouldAddEndPuncttrue
\mciteSetBstMidEndSepPunct{\mcitedefaultmidpunct}
{\mcitedefaultendpunct}{\mcitedefaultseppunct}\relax
\EndOfBibitem
\bibitem[Goodpaster \latin{et~al.}(2014)Goodpaster, Barnes, Manby, and
  Miller]{Goodpaster2014}
Goodpaster,~J.~D.; Barnes,~T.~A.; Manby,~F.~R.; Miller,~T.~F. Accurate and
  systematically improvable density functional theory embedding for correlated
  wavefunctions. \emph{J. Chem. Phys.} \textbf{2014}, \emph{140}, 18A507\relax
\mciteBstWouldAddEndPuncttrue
\mciteSetBstMidEndSepPunct{\mcitedefaultmidpunct}
{\mcitedefaultendpunct}{\mcitedefaultseppunct}\relax
\EndOfBibitem
\bibitem[Jacob and Neugebauer(2014)Jacob, and Neugebauer]{Jacob2014}
Jacob,~C.~R.; Neugebauer,~J. Subsystem density-functional theory. \emph{Wiley
  Interdiscip. Rev.: Comput. Mol. Sci.} \textbf{2014}, \emph{4}, 325--362\relax
\mciteBstWouldAddEndPuncttrue
\mciteSetBstMidEndSepPunct{\mcitedefaultmidpunct}
{\mcitedefaultendpunct}{\mcitedefaultseppunct}\relax
\EndOfBibitem
\bibitem[Genova \latin{et~al.}(2014)Genova, Ceresoli, and
  Pavanello]{Genova2014}
Genova,~A.; Ceresoli,~D.; Pavanello,~M. Periodic subsystem density-functional
  theory. \emph{J. Chem. Phys.} \textbf{2014}, \emph{141}, 174101\relax
\mciteBstWouldAddEndPuncttrue
\mciteSetBstMidEndSepPunct{\mcitedefaultmidpunct}
{\mcitedefaultendpunct}{\mcitedefaultseppunct}\relax
\EndOfBibitem
\bibitem[Wen \latin{et~al.}(2019)Wen, Graham, Chulhai, and Goodpaster]{Wen2019}
Wen,~X.; Graham,~D.~S.; Chulhai,~D.~V.; Goodpaster,~J.~D. Absolutely Localized
  Projection-Based Embedding for Excited States. \emph{J. Chem. Theory Comput.}
  \textbf{2019}, \emph{16}, 385--398\relax
\mciteBstWouldAddEndPuncttrue
\mciteSetBstMidEndSepPunct{\mcitedefaultmidpunct}
{\mcitedefaultendpunct}{\mcitedefaultseppunct}\relax
\EndOfBibitem
\bibitem[Knizia and Chan(2012)Knizia, and Chan]{Knizia2012}
Knizia,~G.; Chan,~G. K.-L. Density Matrix Embedding: A Simple Alternative to
  Dynamical Mean-Field Theory. \emph{Phys. Rev. Lett.} \textbf{2012},
  \emph{109}, 186404\relax
\mciteBstWouldAddEndPuncttrue
\mciteSetBstMidEndSepPunct{\mcitedefaultmidpunct}
{\mcitedefaultendpunct}{\mcitedefaultseppunct}\relax
\EndOfBibitem
\bibitem[Wouters \latin{et~al.}(2016)Wouters, Jim{\'{e}}nez-Hoyos, Sun, and
  Chan]{Wouters2016}
Wouters,~S.; Jim{\'{e}}nez-Hoyos,~C.~A.; Sun,~Q.; Chan,~G. K.-L. A Practical
  Guide to Density Matrix Embedding Theory in Quantum Chemistry. \emph{J. Chem.
  Theory Comput.} \textbf{2016}, \emph{12}, 2706--2719\relax
\mciteBstWouldAddEndPuncttrue
\mciteSetBstMidEndSepPunct{\mcitedefaultmidpunct}
{\mcitedefaultendpunct}{\mcitedefaultseppunct}\relax
\EndOfBibitem
\bibitem[Pham \latin{et~al.}(2019)Pham, Hermes, and
  Gagliardi]{pham2019periodic}
Pham,~H.~Q.; Hermes,~M.~R.; Gagliardi,~L. Periodic Electronic Structure
  Calculations with the Density Matrix Embedding Theory. \emph{J. Chem. Theory
  Comput.} \textbf{2019}, \emph{16}, 130--140\relax
\mciteBstWouldAddEndPuncttrue
\mciteSetBstMidEndSepPunct{\mcitedefaultmidpunct}
{\mcitedefaultendpunct}{\mcitedefaultseppunct}\relax
\EndOfBibitem
\bibitem[Lan \latin{et~al.}(2016)Lan, Kananenka, and Zgid]{Nguyen2016}
Lan,~T.~N.; Kananenka,~A.~A.; Zgid,~D. Rigorous Ab Initio Quantum Embedding for
  Quantum Chemistry Using Green's Function Theory: Screened Interaction,
  Nonlocal Self-Energy Relaxation, Orbital Basis, and Chemical Accuracy.
  \emph{J. Chem. Theory Comput.} \textbf{2016}, \emph{12}, 4856--4870\relax
\mciteBstWouldAddEndPuncttrue
\mciteSetBstMidEndSepPunct{\mcitedefaultmidpunct}
{\mcitedefaultendpunct}{\mcitedefaultseppunct}\relax
\EndOfBibitem
\bibitem[Dvorak and Rinke(2019)Dvorak, and Rinke]{Dvorak2019}
Dvorak,~M.; Rinke,~P. Dynamical configuration interaction: Quantum embedding
  that combines wave functions and Green's functions. \emph{Phys. Rev. B}
  \textbf{2019}, \emph{99}, 115134\relax
\mciteBstWouldAddEndPuncttrue
\mciteSetBstMidEndSepPunct{\mcitedefaultmidpunct}
{\mcitedefaultendpunct}{\mcitedefaultseppunct}\relax
\EndOfBibitem
\bibitem[Zhu \latin{et~al.}(2019)Zhu, Jim{\'e}nez-Hoyos, McClain, Berkelbach,
  and Chan]{Zhu2019}
Zhu,~T.; Jim{\'e}nez-Hoyos,~C.~A.; McClain,~J.; Berkelbach,~T.~C.; Chan,~G.
  K.-L. Coupled-cluster impurity solvers for dynamical mean-field theory.
  \emph{Physical Review B} \textbf{2019}, \emph{100}, 115154\relax
\mciteBstWouldAddEndPuncttrue
\mciteSetBstMidEndSepPunct{\mcitedefaultmidpunct}
{\mcitedefaultendpunct}{\mcitedefaultseppunct}\relax
\EndOfBibitem
\bibitem[Aryasetiawan \latin{et~al.}(2004)Aryasetiawan, Imada, Georges,
  Kotliar, Biermann, and Lichtenstein]{Aryasetiawan2004}
Aryasetiawan,~F.; Imada,~M.; Georges,~A.; Kotliar,~G.; Biermann,~S.;
  Lichtenstein,~A.~I. Frequency-dependent local interactions and low-energy
  effective models from electronic structure calculations. \emph{Phys. Rev. B}
  \textbf{2004}, \emph{70}, 195104\relax
\mciteBstWouldAddEndPuncttrue
\mciteSetBstMidEndSepPunct{\mcitedefaultmidpunct}
{\mcitedefaultendpunct}{\mcitedefaultseppunct}\relax
\EndOfBibitem
\bibitem[Aryasetiawan \latin{et~al.}(2009)Aryasetiawan, Tomczak, Miyake, and
  Sakuma]{Aryasetiawan2009}
Aryasetiawan,~F.; Tomczak,~J.~M.; Miyake,~T.; Sakuma,~R. Downfolded Self-Energy
  of Many-Electron Systems. \emph{Phys. Rev. Lett.} \textbf{2009},
  \emph{102}\relax
\mciteBstWouldAddEndPuncttrue
\mciteSetBstMidEndSepPunct{\mcitedefaultmidpunct}
{\mcitedefaultendpunct}{\mcitedefaultseppunct}\relax
\EndOfBibitem
\bibitem[Miyake \latin{et~al.}(2009)Miyake, Aryasetiawan, and
  Imada]{Miyake2009}
Miyake,~T.; Aryasetiawan,~F.; Imada,~M. Ab initio procedure for constructing
  effective models of correlated materials with entangled band structure.
  \emph{Phys. Rev. B} \textbf{2009}, \emph{80}, 155134\relax
\mciteBstWouldAddEndPuncttrue
\mciteSetBstMidEndSepPunct{\mcitedefaultmidpunct}
{\mcitedefaultendpunct}{\mcitedefaultseppunct}\relax
\EndOfBibitem
\bibitem[Imada and Miyake(2010)Imada, and Miyake]{Imada2010}
Imada,~M.; Miyake,~T. Electronic Structure Calculation by First Principles for
  Strongly Correlated Electron Systems. \emph{J. Phys. Soc. Jpn.}
  \textbf{2010}, \emph{79}, 112001\relax
\mciteBstWouldAddEndPuncttrue
\mciteSetBstMidEndSepPunct{\mcitedefaultmidpunct}
{\mcitedefaultendpunct}{\mcitedefaultseppunct}\relax
\EndOfBibitem
\bibitem[Hirayama \latin{et~al.}(2013)Hirayama, Miyake, and
  Imada]{Hirayama2013}
Hirayama,~M.; Miyake,~T.; Imada,~M. Derivation of static low-energy effective
  models by an ab initio downfolding method without double counting of Coulomb
  correlations: Application to \ce{SrVO_3}, FeSe, and FeTe. \emph{Phys. Rev. B}
  \textbf{2013}, \emph{87}, 195144\relax
\mciteBstWouldAddEndPuncttrue
\mciteSetBstMidEndSepPunct{\mcitedefaultmidpunct}
{\mcitedefaultendpunct}{\mcitedefaultseppunct}\relax
\EndOfBibitem
\bibitem[Hirayama \latin{et~al.}(2017)Hirayama, Miyake, Imada, and
  Biermann]{Hirayama2017}
Hirayama,~M.; Miyake,~T.; Imada,~M.; Biermann,~S. Low-energy effective
  Hamiltonians for correlated electron systems beyond density functional
  theory. \emph{Phys. Rev. B} \textbf{2017}, \emph{96}, 075102\relax
\mciteBstWouldAddEndPuncttrue
\mciteSetBstMidEndSepPunct{\mcitedefaultmidpunct}
{\mcitedefaultendpunct}{\mcitedefaultseppunct}\relax
\EndOfBibitem
\bibitem[Cho and Berkelbach(2018)Cho, and Berkelbach]{Cho2018}
Cho,~Y.; Berkelbach,~T.~C. Environmentally sensitive theory of electronic and
  optical transitions in atomically thin semiconductors. \emph{Phys. Rev. B}
  \textbf{2018}, \emph{97}, 041409\relax
\mciteBstWouldAddEndPuncttrue
\mciteSetBstMidEndSepPunct{\mcitedefaultmidpunct}
{\mcitedefaultendpunct}{\mcitedefaultseppunct}\relax
\EndOfBibitem
\bibitem[Romanova and Vl{\v{c}}ek(2020)Romanova, and Vl{\v{c}}ek]{Romanova2020}
Romanova,~M.; Vl{\v{c}}ek,~V. Decomposition and embedding in the stochastic GW
  self-energy. \emph{J. Chem. Phys.} \textbf{2020}, \emph{153}, 134103\relax
\mciteBstWouldAddEndPuncttrue
\mciteSetBstMidEndSepPunct{\mcitedefaultmidpunct}
{\mcitedefaultendpunct}{\mcitedefaultseppunct}\relax
\EndOfBibitem
\bibitem[Aryasetiawan \latin{et~al.}(2006)Aryasetiawan, Karlsson, Jepsen, and
  Schönberger]{Aryasetiawan2006}
Aryasetiawan,~F.; Karlsson,~K.; Jepsen,~O.; Schönberger,~U. Calculations of
  {Hubbard U from} first-principles. \emph{Phys. Rev. B} \textbf{2006},
  \emph{74}\relax
\mciteBstWouldAddEndPuncttrue
\mciteSetBstMidEndSepPunct{\mcitedefaultmidpunct}
{\mcitedefaultendpunct}{\mcitedefaultseppunct}\relax
\EndOfBibitem
\bibitem[Shih \latin{et~al.}(2012)Shih, Zhang, Zhang, and Zhang]{Shih2012}
Shih,~B.-C.; Zhang,~Y.; Zhang,~W.; Zhang,~P. Screened Coulomb interaction of
  localized electrons in solids from first principles. \emph{Phys. Rev. B}
  \textbf{2012}, \emph{85}\relax
\mciteBstWouldAddEndPuncttrue
\mciteSetBstMidEndSepPunct{\mcitedefaultmidpunct}
{\mcitedefaultendpunct}{\mcitedefaultseppunct}\relax
\EndOfBibitem
\bibitem[Nilsson and Aryasetiawan(2017)Nilsson, and Aryasetiawan]{Nilsson2017}
Nilsson,~F.; Aryasetiawan,~F. Electronic structure of strongly correlated
  materials: from one-particle to many-body theory. \emph{Mater. Res. Express}
  \textbf{2017}, \emph{4}, 034001\relax
\mciteBstWouldAddEndPuncttrue
\mciteSetBstMidEndSepPunct{\mcitedefaultmidpunct}
{\mcitedefaultendpunct}{\mcitedefaultseppunct}\relax
\EndOfBibitem
\bibitem[Tadano \latin{et~al.}(2019)Tadano, Nomura, and Imada]{Tadano2019}
Tadano,~T.; Nomura,~Y.; Imada,~M. Ab initio derivation of an effective
  Hamiltonian for the \ce{La_2 Cu O_4}/\ce{La_{1.55} Sr_{0.45} Cu O_4}
  heterostructure. \emph{Phys. Rev. B} \textbf{2019}, \emph{99}, 155148\relax
\mciteBstWouldAddEndPuncttrue
\mciteSetBstMidEndSepPunct{\mcitedefaultmidpunct}
{\mcitedefaultendpunct}{\mcitedefaultseppunct}\relax
\EndOfBibitem
\bibitem[Wehling \latin{et~al.}(2011)Wehling, \ifmmode \mbox{\c{S}}\else
  \c{S}\fi{}a\ifmmode \mbox{\c{s}}\else \c{s}\fi{}\ifmmode \imath \else \i
  \fi{}o\ifmmode~\breve{g}\else \u{g}\fi{}lu, Friedrich, Lichtenstein,
  Katsnelson, and Bl\"ugel]{Wehling2011}
Wehling,~T.~O.; \ifmmode \mbox{\c{S}}\else \c{S}\fi{}a\ifmmode
  \mbox{\c{s}}\else \c{s}\fi{}\ifmmode \imath \else \i
  \fi{}o\ifmmode~\breve{g}\else \u{g}\fi{}lu,~E.; Friedrich,~C.;
  Lichtenstein,~A.~I.; Katsnelson,~M.~I.; Bl\"ugel,~S. Strength of Effective
  Coulomb Interactions in Graphene and Graphite. \emph{Phys. Rev. Lett.}
  \textbf{2011}, \emph{106}, 236805\relax
\mciteBstWouldAddEndPuncttrue
\mciteSetBstMidEndSepPunct{\mcitedefaultmidpunct}
{\mcitedefaultendpunct}{\mcitedefaultseppunct}\relax
\EndOfBibitem
\bibitem[Honerkamp \latin{et~al.}(2018)Honerkamp, Shinaoka, Assaad, and
  Werner]{Honerkamp2018}
Honerkamp,~C.; Shinaoka,~H.; Assaad,~F.~F.; Werner,~P. Limitations of
  constrained random phase approximation downfolding. \emph{Phys. Rev. B}
  \textbf{2018}, \emph{98}, 235151\relax
\mciteBstWouldAddEndPuncttrue
\mciteSetBstMidEndSepPunct{\mcitedefaultmidpunct}
{\mcitedefaultendpunct}{\mcitedefaultseppunct}\relax
\EndOfBibitem
\bibitem[Adler(1962)]{Adler1962}
Adler,~S.~L. Quantum Theory of the Dielectric Constant in Real Solids.
  \emph{Phys. Rev.} \textbf{1962}, \emph{126}, 413--420\relax
\mciteBstWouldAddEndPuncttrue
\mciteSetBstMidEndSepPunct{\mcitedefaultmidpunct}
{\mcitedefaultendpunct}{\mcitedefaultseppunct}\relax
\EndOfBibitem
\bibitem[Wiser(1963)]{Wiser1963}
Wiser,~N. Dielectric Constant with Local Field Effects Included. \emph{Phys.
  Rev.} \textbf{1963}, \emph{129}, 62--69\relax
\mciteBstWouldAddEndPuncttrue
\mciteSetBstMidEndSepPunct{\mcitedefaultmidpunct}
{\mcitedefaultendpunct}{\mcitedefaultseppunct}\relax
\EndOfBibitem
\bibitem[Ma \latin{et~al.}(2020)Ma, Govoni, and Galli]{Ma2020}
Ma,~H.; Govoni,~M.; Galli,~G. Quantum simulations of materials on near-term
  quantum computers. \emph{npj Comput. Mater.} \textbf{2020}, \emph{6},
  85\relax
\mciteBstWouldAddEndPuncttrue
\mciteSetBstMidEndSepPunct{\mcitedefaultmidpunct}
{\mcitedefaultendpunct}{\mcitedefaultseppunct}\relax
\EndOfBibitem
\bibitem[Ma \latin{et~al.}(2018)Ma, Govoni, Gygi, and Galli]{Ma2018}
Ma,~H.; Govoni,~M.; Gygi,~F.; Galli,~G. A Finite-Field Approach
  {forGWCalculations} beyond the Random Phase Approximation. \emph{J. Chem.
  Theory Comput.} \textbf{2018}, \emph{15}, 154--164\relax
\mciteBstWouldAddEndPuncttrue
\mciteSetBstMidEndSepPunct{\mcitedefaultmidpunct}
{\mcitedefaultendpunct}{\mcitedefaultseppunct}\relax
\EndOfBibitem
\bibitem[Nguyen \latin{et~al.}(2019)Nguyen, Ma, Govoni, Gygi, and
  Galli]{Nguyen2019}
Nguyen,~N.~L.; Ma,~H.; Govoni,~M.; Gygi,~F.; Galli,~G. Finite-Field Approach to
  Solving the Bethe-Salpeter Equation. \emph{Phys. Rev. Lett.} \textbf{2019},
  \emph{122}, 237402\relax
\mciteBstWouldAddEndPuncttrue
\mciteSetBstMidEndSepPunct{\mcitedefaultmidpunct}
{\mcitedefaultendpunct}{\mcitedefaultseppunct}\relax
\EndOfBibitem
\bibitem[Wilson \latin{et~al.}(2008)Wilson, Gygi, and Galli]{Wilson2008}
Wilson,~H.~F.; Gygi,~F.; Galli,~G. Efficient iterative method for calculations
  of dielectric matrices. \emph{Phys. Rev. B} \textbf{2008}, \emph{78},
  113303\relax
\mciteBstWouldAddEndPuncttrue
\mciteSetBstMidEndSepPunct{\mcitedefaultmidpunct}
{\mcitedefaultendpunct}{\mcitedefaultseppunct}\relax
\EndOfBibitem
\bibitem[Nguyen \latin{et~al.}(2012)Nguyen, Pham, Rocca, and Galli]{Nguyen2012}
Nguyen,~H.-V.; Pham,~T.~A.; Rocca,~D.; Galli,~G. Improving accuracy and
  efficiency of calculations of photoemission spectra within the many-body
  perturbation theory. \emph{Phys. Rev. B} \textbf{2012}, \emph{85},
  081101\relax
\mciteBstWouldAddEndPuncttrue
\mciteSetBstMidEndSepPunct{\mcitedefaultmidpunct}
{\mcitedefaultendpunct}{\mcitedefaultseppunct}\relax
\EndOfBibitem
\bibitem[Pham \latin{et~al.}(2013)Pham, Nguyen, Rocca, and Galli]{Pham2013}
Pham,~T.~A.; Nguyen,~H.-V.; Rocca,~D.; Galli,~G. $GW$ calculations using the
  spectral decomposition of the dielectric matrix: Verification, validation,
  and comparison of methods. \emph{Phys. Rev. B} \textbf{2013}, \emph{87},
  155148\relax
\mciteBstWouldAddEndPuncttrue
\mciteSetBstMidEndSepPunct{\mcitedefaultmidpunct}
{\mcitedefaultendpunct}{\mcitedefaultseppunct}\relax
\EndOfBibitem
\bibitem[Govoni and Galli(2015)Govoni, and Galli]{Govoni2015}
Govoni,~M.; Galli,~G. Large Scale {GW} Calculations. \emph{J. Chem. Theory
  Comput.} \textbf{2015}, \emph{11}, 2680--2696\relax
\mciteBstWouldAddEndPuncttrue
\mciteSetBstMidEndSepPunct{\mcitedefaultmidpunct}
{\mcitedefaultendpunct}{\mcitedefaultseppunct}\relax
\EndOfBibitem
\bibitem[Ma \latin{et~al.}(2020)Ma, Sheng, Govoni, and Galli]{Ma2020pccp}
Ma,~H.; Sheng,~N.; Govoni,~M.; Galli,~G. First-principles studies of strongly
  correlated states in defect spin qubits in diamond. \emph{Phys. Chem. Chem.
  Phys.} \textbf{2020}, \relax
\mciteBstWouldAddEndPunctfalse
\mciteSetBstMidEndSepPunct{\mcitedefaultmidpunct}
{}{\mcitedefaultseppunct}\relax
\EndOfBibitem
\bibitem[Marzari \latin{et~al.}(2012)Marzari, Mostofi, Yates, Souza, and
  Vanderbilt]{marzari2012maximally}
Marzari,~N.; Mostofi,~A.~A.; Yates,~J.~R.; Souza,~I.; Vanderbilt,~D. Maximally
  localized Wannier functions: Theory and applications. \emph{Rev. Mod. Phys.}
  \textbf{2012}, \emph{84}, 1419\relax
\mciteBstWouldAddEndPuncttrue
\mciteSetBstMidEndSepPunct{\mcitedefaultmidpunct}
{\mcitedefaultendpunct}{\mcitedefaultseppunct}\relax
\EndOfBibitem
\bibitem[Timrov \latin{et~al.}(2018)Timrov, Marzari, and
  Cococcioni]{Timrov2018}
Timrov,~I.; Marzari,~N.; Cococcioni,~M. Hubbard parameters from
  density-functional perturbation theory. \emph{Phys. Rev. B} \textbf{2018},
  \emph{98}, 085127\relax
\mciteBstWouldAddEndPuncttrue
\mciteSetBstMidEndSepPunct{\mcitedefaultmidpunct}
{\mcitedefaultendpunct}{\mcitedefaultseppunct}\relax
\EndOfBibitem
\bibitem[Hedin(1965)]{Hedin1965}
Hedin,~L. New Method for Calculating the One-Particle Green's Function with
  Application to the Electron-Gas Problem. \emph{Phys. Rev.} \textbf{1965},
  \emph{139}, A796--A823\relax
\mciteBstWouldAddEndPuncttrue
\mciteSetBstMidEndSepPunct{\mcitedefaultmidpunct}
{\mcitedefaultendpunct}{\mcitedefaultseppunct}\relax
\EndOfBibitem
\bibitem[Not()]{Note-1}
Some literatures define $\chi_0^\text{A}$ by adding $O^\text{A}$ to all the
  four appearances of Kohn-Sham orbitals in Eq. \ref{chi0A}. This definition is
  equivalent to our definition if the active space is spanned by a set of
  Kohn-Sham orbitals. For general active spaces, we tested quantum embedding
  calculations using both definitions and we found the difference in results
  (e.g. Hubbard parameters of \ce{SrTiO3}) is negligible.\relax
\mciteBstWouldAddEndPunctfalse
\mciteSetBstMidEndSepPunct{\mcitedefaultmidpunct}
{}{\mcitedefaultseppunct}\relax
\EndOfBibitem
\bibitem[Liechtenstein \latin{et~al.}(1995)Liechtenstein, Anisimov, and
  Zaanen]{Liechtenstein1995}
Liechtenstein,~A.~I.; Anisimov,~V.~I.; Zaanen,~J. Density-functional theory and
  strong interactions: Orbital ordering in Mott-Hubbard insulators. \emph{Phys.
  Rev. B} \textbf{1995}, \emph{52}, R5467--R5470\relax
\mciteBstWouldAddEndPuncttrue
\mciteSetBstMidEndSepPunct{\mcitedefaultmidpunct}
{\mcitedefaultendpunct}{\mcitedefaultseppunct}\relax
\EndOfBibitem
\bibitem[Ryee and Han(2018)Ryee, and Han]{Ryee2018}
Ryee,~S.; Han,~M.~J. The effect of double counting, spin density, and Hund
  interaction in the different DFT+ U functionals. \emph{Sci. Rep.}
  \textbf{2018}, \emph{8}, 1--11\relax
\mciteBstWouldAddEndPuncttrue
\mciteSetBstMidEndSepPunct{\mcitedefaultmidpunct}
{\mcitedefaultendpunct}{\mcitedefaultseppunct}\relax
\EndOfBibitem
\bibitem[Not()]{Note-2}
In principle, one can also define an electron-test-charge screened Coulomb
  interaction $W_{\text{etc}}$ that represents the screened interaction between
  a test charge and an electron. However, it is difficult to apply the
  cRPA-type treatment to $W_{\text{etc}}$ and define a partially screened
  interaction because it is difficult to write $W_{\text{etc}}$ in the form of
  a Dyson-like equation, so we will not consider $W_{\text{etc}}$ in this
  work.\relax
\mciteBstWouldAddEndPunctfalse
\mciteSetBstMidEndSepPunct{\mcitedefaultmidpunct}
{}{\mcitedefaultseppunct}\relax
\EndOfBibitem
\bibitem[Hybertsen and Louie(1986)Hybertsen, and Louie]{Hybertson1986}
Hybertsen,~M.~S.; Louie,~S.~G. Electron correlation in semiconductors and
  insulators: Band gaps and quasiparticle energies. \emph{Phys. Rev. B}
  \textbf{1986}, \emph{34}, 5390\relax
\mciteBstWouldAddEndPuncttrue
\mciteSetBstMidEndSepPunct{\mcitedefaultmidpunct}
{\mcitedefaultendpunct}{\mcitedefaultseppunct}\relax
\EndOfBibitem
\bibitem[Sole \latin{et~al.}(1994)Sole, Reining, and Godby]{DelSole1994}
Sole,~R.~D.; Reining,~L.; Godby,~R.~W. GW$\Gamma$ approximation for electron
  self-energies in semiconductors and insulators. \emph{Phys. Rev. B}
  \textbf{1994}, \emph{49}, 8024--8028\relax
\mciteBstWouldAddEndPuncttrue
\mciteSetBstMidEndSepPunct{\mcitedefaultmidpunct}
{\mcitedefaultendpunct}{\mcitedefaultseppunct}\relax
\EndOfBibitem
\bibitem[Martin \latin{et~al.}(2016)Martin, Reining, and Ceperley]{Martin2016}
Martin,~R.~M.; Reining,~L.; Ceperley,~D.~M. \emph{Interacting Electrons: Theory
  and Computational Approaches}; Cambridge University Press, 2016\relax
\mciteBstWouldAddEndPuncttrue
\mciteSetBstMidEndSepPunct{\mcitedefaultmidpunct}
{\mcitedefaultendpunct}{\mcitedefaultseppunct}\relax
\EndOfBibitem
\bibitem[Paier \latin{et~al.}(2008)Paier, Marsman, and Kresse]{Paier2008}
Paier,~J.; Marsman,~M.; Kresse,~G. Dielectric properties and excitons for
  extended systems from hybrid functionals. \emph{Phys. Rev. B} \textbf{2008},
  \emph{78}, 121201\relax
\mciteBstWouldAddEndPuncttrue
\mciteSetBstMidEndSepPunct{\mcitedefaultmidpunct}
{\mcitedefaultendpunct}{\mcitedefaultseppunct}\relax
\EndOfBibitem
\bibitem[Gr{\"u}neis \latin{et~al.}(2014)Gr{\"u}neis, Kresse, Hinuma, and
  Oba]{Gruneis2014}
Gr{\"u}neis,~A.; Kresse,~G.; Hinuma,~Y.; Oba,~F. Ionization Potentials of
  Solids: The Importance of Vertex Corrections. \emph{Phys. Rev. Lett.}
  \textbf{2014}, \emph{112}, 096401\relax
\mciteBstWouldAddEndPuncttrue
\mciteSetBstMidEndSepPunct{\mcitedefaultmidpunct}
{\mcitedefaultendpunct}{\mcitedefaultseppunct}\relax
\EndOfBibitem
\bibitem[McAvoy \latin{et~al.}(2018)McAvoy, Govoni, and Galli]{McAvoy2018}
McAvoy,~R.~L.; Govoni,~M.; Galli,~G. Coupling First-Principles Calculations of
  Electron{\textendash}Electron and Electron{\textendash}Phonon Scattering, and
  Applications to Carbon-Based Nanostructures. \emph{J. Chem. Theory Comput.}
  \textbf{2018}, \emph{14}, 6269--6275\relax
\mciteBstWouldAddEndPuncttrue
\mciteSetBstMidEndSepPunct{\mcitedefaultmidpunct}
{\mcitedefaultendpunct}{\mcitedefaultseppunct}\relax
\EndOfBibitem
\bibitem[Giustino(2017)]{Giustino2017}
Giustino,~F. Electron-phonon interactions from first principles. \emph{Rev.
  Mod. Phys.} \textbf{2017}, \emph{89}, 015003\relax
\mciteBstWouldAddEndPuncttrue
\mciteSetBstMidEndSepPunct{\mcitedefaultmidpunct}
{\mcitedefaultendpunct}{\mcitedefaultseppunct}\relax
\EndOfBibitem
\bibitem[Gygi and Baldereschi(1986)Gygi, and Baldereschi]{Gygi1986}
Gygi,~F.; Baldereschi,~A. Self-consistent Hartree-Fock and screened-exchange
  calculations in solids: Application to silicon. \emph{Phys. Rev. B}
  \textbf{1986}, \emph{34}, 4405--4408\relax
\mciteBstWouldAddEndPuncttrue
\mciteSetBstMidEndSepPunct{\mcitedefaultmidpunct}
{\mcitedefaultendpunct}{\mcitedefaultseppunct}\relax
\EndOfBibitem
\bibitem[Gygi(2008)]{Gygi2008}
Gygi,~F. Architecture of Qbox: A scalable first-principles molecular dynamics
  code. \emph{IBM J. Res. Dev.} \textbf{2008}, \emph{52}, 137--144\relax
\mciteBstWouldAddEndPuncttrue
\mciteSetBstMidEndSepPunct{\mcitedefaultmidpunct}
{\mcitedefaultendpunct}{\mcitedefaultseppunct}\relax
\EndOfBibitem
\bibitem[Giannozzi \latin{et~al.}(2009)Giannozzi, Baroni, Bonini, Calandra,
  Car, Cavazzoni, Ceresoli, Chiarotti, Cococcioni, Dabo, Corso, de~Gironcoli,
  Fabris, Fratesi, Gebauer, Gerstmann, Gougoussis, Kokalj, Lazzeri,
  Martin-Samos, Marzari, Mauri, Mazzarello, Paolini, Pasquarello, Paulatto,
  Sbraccia, Scandolo, Sclauzero, Seitsonen, Smogunov, Umari, and
  Wentzcovitch]{Giannozzi2009}
Giannozzi,~P.; Baroni,~S.; Bonini,~N.; Calandra,~M.; Car,~R.; Cavazzoni,~C.;
  Ceresoli,~D.; Chiarotti,~G.~L.; Cococcioni,~M.; Dabo,~I.; Corso,~A.~D.;
  de~Gironcoli,~S.; Fabris,~S.; Fratesi,~G.; Gebauer,~R.; Gerstmann,~U.;
  Gougoussis,~C.; Kokalj,~A.; Lazzeri,~M.; Martin-Samos,~L.; Marzari,~N.;
  Mauri,~F.; Mazzarello,~R.; Paolini,~S.; Pasquarello,~A.; Paulatto,~L.;
  Sbraccia,~C.; Scandolo,~S.; Sclauzero,~G.; Seitsonen,~A.~P.; Smogunov,~A.;
  Umari,~P.; Wentzcovitch,~R.~M. {QUANTUM} {ESPRESSO}: a modular and
  open-source software project for quantum simulations of materials. \emph{J.
  Phys.: Condens. Matter} \textbf{2009}, \emph{21}, 395502\relax
\mciteBstWouldAddEndPuncttrue
\mciteSetBstMidEndSepPunct{\mcitedefaultmidpunct}
{\mcitedefaultendpunct}{\mcitedefaultseppunct}\relax
\EndOfBibitem
\bibitem[Mostofi \latin{et~al.}(2008)Mostofi, Yates, Lee, Souza, Vanderbilt,
  and Marzari]{mostofi2008wannier90}
Mostofi,~A.~A.; Yates,~J.~R.; Lee,~Y.-S.; Souza,~I.; Vanderbilt,~D.;
  Marzari,~N. wannier90: A tool for obtaining maximally-localised Wannier
  functions. \emph{Comput. Phys. Commun.} \textbf{2008}, \emph{178},
  685--699\relax
\mciteBstWouldAddEndPuncttrue
\mciteSetBstMidEndSepPunct{\mcitedefaultmidpunct}
{\mcitedefaultendpunct}{\mcitedefaultseppunct}\relax
\EndOfBibitem
\bibitem[Sun \latin{et~al.}(2017)Sun, Berkelbach, Blunt, Booth, Guo, Li, Liu,
  McClain, Sayfutyarova, Sharma, Wouters, and Chan]{Sun2017}
Sun,~Q.; Berkelbach,~T.~C.; Blunt,~N.~S.; Booth,~G.~H.; Guo,~S.; Li,~Z.;
  Liu,~J.; McClain,~J.~D.; Sayfutyarova,~E.~R.; Sharma,~S.; Wouters,~S.;
  Chan,~G. K.-L. Py {SCF}: the Python-based simulations of chemistry framework.
  \emph{Wiley Interdiscip. Rev.: Comput. Mol. Sci.} \textbf{2017}, \emph{8},
  e1340\relax
\mciteBstWouldAddEndPuncttrue
\mciteSetBstMidEndSepPunct{\mcitedefaultmidpunct}
{\mcitedefaultendpunct}{\mcitedefaultseppunct}\relax
\EndOfBibitem
\bibitem[Schlipf and Gygi(2015)Schlipf, and Gygi]{Schlipf2015}
Schlipf,~M.; Gygi,~F. Optimization algorithm for the generation of {ONCV}
  pseudopotentials. \emph{Comput. Phys. Commun.} \textbf{2015}, \emph{196},
  36--44\relax
\mciteBstWouldAddEndPuncttrue
\mciteSetBstMidEndSepPunct{\mcitedefaultmidpunct}
{\mcitedefaultendpunct}{\mcitedefaultseppunct}\relax
\EndOfBibitem
\bibitem[Weber \latin{et~al.}(2010)Weber, Koehl, Varley, Janotti, Buckley,
  de~Walle, and Awschalom]{Weber2010}
Weber,~J.~R.; Koehl,~W.~F.; Varley,~J.~B.; Janotti,~A.; Buckley,~B.~B.;
  de~Walle,~C. G.~V.; Awschalom,~D.~D. Quantum computing with defects.
  \emph{Proc. Natl. Acad. Sci. U. S. A.} \textbf{2010}, \emph{107},
  8513--8518\relax
\mciteBstWouldAddEndPuncttrue
\mciteSetBstMidEndSepPunct{\mcitedefaultmidpunct}
{\mcitedefaultendpunct}{\mcitedefaultseppunct}\relax
\EndOfBibitem
\bibitem[Seo \latin{et~al.}(2016)Seo, Govoni, and Galli]{Seo2016}
Seo,~H.; Govoni,~M.; Galli,~G. Design of defect spins in piezoelectric aluminum
  nitride for solid-state hybrid quantum technologies. \emph{Sci. Rep.}
  \textbf{2016}, \emph{6}, 20803\relax
\mciteBstWouldAddEndPuncttrue
\mciteSetBstMidEndSepPunct{\mcitedefaultmidpunct}
{\mcitedefaultendpunct}{\mcitedefaultseppunct}\relax
\EndOfBibitem
\bibitem[Seo \latin{et~al.}(2017)Seo, Ma, Govoni, and Galli]{Seo2017}
Seo,~H.; Ma,~H.; Govoni,~M.; Galli,~G. Designing defect-based qubit candidates
  in wide-gap binary semiconductors for solid-state quantum technologies.
  \emph{Phys. Rev. Mater.} \textbf{2017}, \emph{1}, 075002\relax
\mciteBstWouldAddEndPuncttrue
\mciteSetBstMidEndSepPunct{\mcitedefaultmidpunct}
{\mcitedefaultendpunct}{\mcitedefaultseppunct}\relax
\EndOfBibitem
\bibitem[Iv{\'{a}}dy \latin{et~al.}(2018)Iv{\'{a}}dy, Abrikosov, and
  Gali]{Ivady2018}
Iv{\'{a}}dy,~V.; Abrikosov,~I.~A.; Gali,~A. First principles calculation of
  spin-related quantities for point defect qubit research. \emph{npj Comput.
  Mater.} \textbf{2018}, \emph{4}\relax
\mciteBstWouldAddEndPuncttrue
\mciteSetBstMidEndSepPunct{\mcitedefaultmidpunct}
{\mcitedefaultendpunct}{\mcitedefaultseppunct}\relax
\EndOfBibitem
\bibitem[Dreyer \latin{et~al.}(2018)Dreyer, Alkauskas, Lyons, Janotti, and
  Van~de Walle]{Dreyer2018}
Dreyer,~C.~E.; Alkauskas,~A.; Lyons,~J.~L.; Janotti,~A.; Van~de Walle,~C.~G.
  First-Principles Calculations of Point Defects for Quantum Technologies.
  \emph{Annu. Rev. Mater. Res.} \textbf{2018}, \emph{48}, 1--26\relax
\mciteBstWouldAddEndPuncttrue
\mciteSetBstMidEndSepPunct{\mcitedefaultmidpunct}
{\mcitedefaultendpunct}{\mcitedefaultseppunct}\relax
\EndOfBibitem
\bibitem[Anderson \latin{et~al.}(2019)Anderson, Bourassa, Miao, Wolfowicz,
  Mintun, Crook, Abe, Ul~Hassan, Son, Ohshima, and Awschalom]{Anderson2019}
Anderson,~C.~P.; Bourassa,~A.; Miao,~K.~C.; Wolfowicz,~G.; Mintun,~P.~J.;
  Crook,~A.~L.; Abe,~H.; Ul~Hassan,~J.; Son,~N.~T.; Ohshima,~T.;
  Awschalom,~D.~D. Electrical and optical control of single spins integrated in
  scalable semiconductor devices. \emph{Science} \textbf{2019}, \emph{366},
  1225--1230\relax
\mciteBstWouldAddEndPuncttrue
\mciteSetBstMidEndSepPunct{\mcitedefaultmidpunct}
{\mcitedefaultendpunct}{\mcitedefaultseppunct}\relax
\EndOfBibitem
\bibitem[Maze \latin{et~al.}(2011)Maze, Gali, Togan, Chu, Trifonov, Kaxiras,
  and Lukin]{Maze2011}
Maze,~J.~R.; Gali,~A.; Togan,~E.; Chu,~Y.; Trifonov,~A.; Kaxiras,~E.;
  Lukin,~M.~D. Properties of nitrogen-vacancy centers in diamond: the group
  theoretic approach. \emph{New J. Phys.} \textbf{2011}, \emph{13},
  025025\relax
\mciteBstWouldAddEndPuncttrue
\mciteSetBstMidEndSepPunct{\mcitedefaultmidpunct}
{\mcitedefaultendpunct}{\mcitedefaultseppunct}\relax
\EndOfBibitem
\bibitem[Doherty \latin{et~al.}(2011)Doherty, Manson, Delaney, and
  Hollenberg]{Doherty2011}
Doherty,~M.~W.; Manson,~N.~B.; Delaney,~P.; Hollenberg,~L. C.~L. The negatively
  charged nitrogen-vacancy centre in diamond: the electronic solution.
  \emph{New J. Phys.} \textbf{2011}, \emph{13}, 025019\relax
\mciteBstWouldAddEndPuncttrue
\mciteSetBstMidEndSepPunct{\mcitedefaultmidpunct}
{\mcitedefaultendpunct}{\mcitedefaultseppunct}\relax
\EndOfBibitem
\bibitem[Perdew \latin{et~al.}(1996)Perdew, Burke, and Ernzerhof]{Perdew1996}
Perdew,~J.~P.; Burke,~K.; Ernzerhof,~M. Generalized Gradient Approximation Made
  Simple. \emph{Phys. Rev. Lett.} \textbf{1996}, \emph{77}, 3865--3868\relax
\mciteBstWouldAddEndPuncttrue
\mciteSetBstMidEndSepPunct{\mcitedefaultmidpunct}
{\mcitedefaultendpunct}{\mcitedefaultseppunct}\relax
\EndOfBibitem
\bibitem[Skone \latin{et~al.}(2014)Skone, Govoni, and Galli]{Skone2014}
Skone,~J.~H.; Govoni,~M.; Galli,~G. Self-consistent hybrid functional for
  condensed systems. \emph{Phys. Rev. B} \textbf{2014}, \emph{89}, 195112\relax
\mciteBstWouldAddEndPuncttrue
\mciteSetBstMidEndSepPunct{\mcitedefaultmidpunct}
{\mcitedefaultendpunct}{\mcitedefaultseppunct}\relax
\EndOfBibitem
\bibitem[Skone \latin{et~al.}(2016)Skone, Govoni, and Galli]{Skone2016}
Skone,~J.~H.; Govoni,~M.; Galli,~G. Nonempirical range-separated hybrid
  functionals for solids and molecules. \emph{Phys. Rev. B} \textbf{2016},
  \emph{93}, 235106\relax
\mciteBstWouldAddEndPuncttrue
\mciteSetBstMidEndSepPunct{\mcitedefaultmidpunct}
{\mcitedefaultendpunct}{\mcitedefaultseppunct}\relax
\EndOfBibitem
\bibitem[Brawand \latin{et~al.}(2016)Brawand, V\"or\"os, Govoni, and
  Galli]{Brawand2016}
Brawand,~N.~P.; V\"or\"os,~M.; Govoni,~M.; Galli,~G. Generalization of
  Dielectric-Dependent Hybrid Functionals to Finite Systems. \emph{Phys. Rev.
  X} \textbf{2016}, \emph{6}, 041002\relax
\mciteBstWouldAddEndPuncttrue
\mciteSetBstMidEndSepPunct{\mcitedefaultmidpunct}
{\mcitedefaultendpunct}{\mcitedefaultseppunct}\relax
\EndOfBibitem
\bibitem[Brawand \latin{et~al.}(2017)Brawand, Govoni, Vörös, and
  Galli]{Brawand2017}
Brawand,~N.~P.; Govoni,~M.; Vörös,~M.; Galli,~G. Performance and
  Self-Consistency of the Generalized Dielectric Dependent Hybrid Functional.
  \emph{J. Chem. Theory Comput.} \textbf{2017}, \emph{13}, 3318--3325, PMID:
  28537727\relax
\mciteBstWouldAddEndPuncttrue
\mciteSetBstMidEndSepPunct{\mcitedefaultmidpunct}
{\mcitedefaultendpunct}{\mcitedefaultseppunct}\relax
\EndOfBibitem
\bibitem[Gerosa \latin{et~al.}(2017)Gerosa, Bottani, Valentin, Onida, and
  Pacchioni]{Gerosa_2017}
Gerosa,~M.; Bottani,~C.~E.; Valentin,~C.~D.; Onida,~G.; Pacchioni,~G. Accuracy
  of dielectric-dependent hybrid functionals in the prediction of
  optoelectronic properties of metal oxide semiconductors: a comprehensive
  comparison with many-body GW and experiments. \emph{J. Phys.: Condens.
  Matter} \textbf{2017}, \emph{30}, 044003\relax
\mciteBstWouldAddEndPuncttrue
\mciteSetBstMidEndSepPunct{\mcitedefaultmidpunct}
{\mcitedefaultendpunct}{\mcitedefaultseppunct}\relax
\EndOfBibitem
\bibitem[Zheng \latin{et~al.}(2019)Zheng, Govoni, and Galli]{Zheng2019}
Zheng,~H.; Govoni,~M.; Galli,~G. Dielectric-dependent hybrid functionals for
  heterogeneous materials. \emph{Phys. Rev. Materials} \textbf{2019}, \emph{3},
  073803\relax
\mciteBstWouldAddEndPuncttrue
\mciteSetBstMidEndSepPunct{\mcitedefaultmidpunct}
{\mcitedefaultendpunct}{\mcitedefaultseppunct}\relax
\EndOfBibitem
\bibitem[Cardona and Peter(2005)Cardona, and Peter]{Cardona2005}
Cardona,~M.; Peter,~Y.~Y. \emph{Fundamentals of semiconductors}; Springer,
  2005\relax
\mciteBstWouldAddEndPuncttrue
\mciteSetBstMidEndSepPunct{\mcitedefaultmidpunct}
{\mcitedefaultendpunct}{\mcitedefaultseppunct}\relax
\EndOfBibitem
\bibitem[Iv{\'a}dy \latin{et~al.}(2020)Iv{\'a}dy, Barcza, Thiering, Li, Hamdi,
  Chou, Legeza, and Gali]{Ivady2020}
Iv{\'a}dy,~V.; Barcza,~G.; Thiering,~G.; Li,~S.; Hamdi,~H.; Chou,~J.-P.;
  Legeza,~{\"O}.; Gali,~A. Ab initio theory of the negatively charged boron
  vacancy qubit in hexagonal boron nitride. \emph{Npj Comput. Mater.}
  \textbf{2020}, \emph{6}, 1--6\relax
\mciteBstWouldAddEndPuncttrue
\mciteSetBstMidEndSepPunct{\mcitedefaultmidpunct}
{\mcitedefaultendpunct}{\mcitedefaultseppunct}\relax
\EndOfBibitem
\bibitem[Knowles and Handy(1984)Knowles, and Handy]{Knowles1984}
Knowles,~P.; Handy,~N. A new determinant-based full configuration interaction
  method. \emph{Chem. Phys. Lett.} \textbf{1984}, \emph{111}, 315--321\relax
\mciteBstWouldAddEndPuncttrue
\mciteSetBstMidEndSepPunct{\mcitedefaultmidpunct}
{\mcitedefaultendpunct}{\mcitedefaultseppunct}\relax
\EndOfBibitem
\bibitem[Davies and Hamer(1976)Davies, and Hamer]{Davies1976}
Davies,~G.; Hamer,~M.~F. Optical Studies of the 1.945 {eV} Vibronic Band in
  Diamond. \emph{Proc. R. Soc. A} \textbf{1976}, \emph{348}, 285--298\relax
\mciteBstWouldAddEndPuncttrue
\mciteSetBstMidEndSepPunct{\mcitedefaultmidpunct}
{\mcitedefaultendpunct}{\mcitedefaultseppunct}\relax
\EndOfBibitem
\bibitem[Rogers \latin{et~al.}(2008)Rogers, Armstrong, Sellars, and
  Manson]{Rogers2008}
Rogers,~L.~J.; Armstrong,~S.; Sellars,~M.~J.; Manson,~N.~B. Infrared emission
  of the {NV} centre in diamond: Zeeman and uniaxial stress studies. \emph{New
  J. Phys.} \textbf{2008}, \emph{10}, 103024\relax
\mciteBstWouldAddEndPuncttrue
\mciteSetBstMidEndSepPunct{\mcitedefaultmidpunct}
{\mcitedefaultendpunct}{\mcitedefaultseppunct}\relax
\EndOfBibitem
\bibitem[Goldman \latin{et~al.}(2015)Goldman, Doherty, Sipahigil, Yao, Bennett,
  Manson, Kubanek, and Lukin]{Goldman2015}
Goldman,~M.~L.; Doherty,~M.~W.; Sipahigil,~A.; Yao,~N.~Y.; Bennett,~S.~D.;
  Manson,~N.~B.; Kubanek,~A.; Lukin,~M.~D. State-selective intersystem crossing
  in nitrogen-vacancy centers. \emph{Phys. Rev. B} \textbf{2015}, \emph{91},
  165201\relax
\mciteBstWouldAddEndPuncttrue
\mciteSetBstMidEndSepPunct{\mcitedefaultmidpunct}
{\mcitedefaultendpunct}{\mcitedefaultseppunct}\relax
\EndOfBibitem
\bibitem[Thiering and Gali(2019)Thiering, and Gali]{Thiering2019}
Thiering,~G.; Gali,~A. The (eg $\otimes$ eu) $\otimes$ Eg product
  Jahn{\textendash}Teller effect in the neutral group-{IV} vacancy quantum bits
  in diamond. \emph{npj Comput. Mater.} \textbf{2019}, \emph{5}, 18\relax
\mciteBstWouldAddEndPuncttrue
\mciteSetBstMidEndSepPunct{\mcitedefaultmidpunct}
{\mcitedefaultendpunct}{\mcitedefaultseppunct}\relax
\EndOfBibitem
\bibitem[Green \latin{et~al.}(2019)Green, Doherty, Nako, Manson,
  D$^\prime$Haenens-Johansson, Williams, Twitchen, and Newton]{Green2019}
Green,~B.~L.; Doherty,~M.~W.; Nako,~E.; Manson,~N.~B.;
  D$^\prime$Haenens-Johansson,~U. F.~S.; Williams,~S.~D.; Twitchen,~D.~J.;
  Newton,~M.~E. Electronic structure of the neutral silicon-vacancy center in
  diamond. \emph{Phys. Rev. B} \textbf{2019}, \emph{99}, 161112\relax
\mciteBstWouldAddEndPuncttrue
\mciteSetBstMidEndSepPunct{\mcitedefaultmidpunct}
{\mcitedefaultendpunct}{\mcitedefaultseppunct}\relax
\EndOfBibitem
\bibitem[Koehl \latin{et~al.}(2011)Koehl, Buckley, Heremans, Calusine, and
  Awschalom]{Koehl2011}
Koehl,~W.~F.; Buckley,~B.~B.; Heremans,~F.~J.; Calusine,~G.; Awschalom,~D.~D.
  Room temperature coherent control of defect spin qubits in silicon carbide.
  \emph{Nature} \textbf{2011}, \emph{479}, 84--87\relax
\mciteBstWouldAddEndPuncttrue
\mciteSetBstMidEndSepPunct{\mcitedefaultmidpunct}
{\mcitedefaultendpunct}{\mcitedefaultseppunct}\relax
\EndOfBibitem
\bibitem[Son \latin{et~al.}(1999)Son, Ellison, Magnusson, MacMillan, Chen,
  Monemar, and Janz{\'{e}}n]{Son1999}
Son,~N.~T.; Ellison,~A.; Magnusson,~B.; MacMillan,~M.~F.; Chen,~W.~M.;
  Monemar,~B.; Janz{\'{e}}n,~E. Photoluminescence and Zeeman effect in
  chromium-doped 4H and 6H {SiC}. \emph{J. Appl. Phys.} \textbf{1999},
  \emph{86}, 4348--4353\relax
\mciteBstWouldAddEndPuncttrue
\mciteSetBstMidEndSepPunct{\mcitedefaultmidpunct}
{\mcitedefaultendpunct}{\mcitedefaultseppunct}\relax
\EndOfBibitem
\bibitem[Solovyev and Imada(2005)Solovyev, and Imada]{Solovyev2005}
Solovyev,~I.~V.; Imada,~M. Screening of Coulomb interactions in transition
  metals. \emph{Phys. Rev. B} \textbf{2005}, \emph{71}\relax
\mciteBstWouldAddEndPuncttrue
\mciteSetBstMidEndSepPunct{\mcitedefaultmidpunct}
{\mcitedefaultendpunct}{\mcitedefaultseppunct}\relax
\EndOfBibitem
\bibitem[Karlsson \latin{et~al.}(2010)Karlsson, Aryasetiawan, and
  Jepsen]{Karlsson2010}
Karlsson,~K.; Aryasetiawan,~F.; Jepsen,~O. Method for calculating the
  electronic structure of correlated materials from a truly first-principles
  LDA+ U scheme. \emph{Phys. Rev. B} \textbf{2010}, \emph{81}, 245113\relax
\mciteBstWouldAddEndPuncttrue
\mciteSetBstMidEndSepPunct{\mcitedefaultmidpunct}
{\mcitedefaultendpunct}{\mcitedefaultseppunct}\relax
\EndOfBibitem
\bibitem[Vaugier \latin{et~al.}(2012)Vaugier, Jiang, and
  Biermann]{vaugier2012hubbard}
Vaugier,~L.; Jiang,~H.; Biermann,~S. Hubbard U and Hund exchange J in
  transition metal oxides: Screening versus localization trends from
  constrained random phase approximation. \emph{Phys. Rev. B} \textbf{2012},
  \emph{86}, 165105\relax
\mciteBstWouldAddEndPuncttrue
\mciteSetBstMidEndSepPunct{\mcitedefaultmidpunct}
{\mcitedefaultendpunct}{\mcitedefaultseppunct}\relax
\EndOfBibitem
\bibitem[Hirayama \latin{et~al.}(2012)Hirayama, Miyake, and
  Imada]{hirayama2012ab}
Hirayama,~M.; Miyake,~T.; Imada,~M. Ab initio Low-energy model of
  transition-metal-oxide heterostructure \ce{LaAlO_3}/\ce{SrTiO_3}. \emph{J.
  Phys. Soc. Jpn.} \textbf{2012}, \emph{81}, 084708\relax
\mciteBstWouldAddEndPuncttrue
\mciteSetBstMidEndSepPunct{\mcitedefaultmidpunct}
{\mcitedefaultendpunct}{\mcitedefaultseppunct}\relax
\EndOfBibitem
\bibitem[Ricca \latin{et~al.}(2020)Ricca, Timrov, Cococcioni, Marzari, and
  Aschauer]{Ricca2020}
Ricca,~C.; Timrov,~I.; Cococcioni,~M.; Marzari,~N.; Aschauer,~U.
  Self-consistent $\mathrm{DFT}+U+V$ study of oxygen vacancies in
  ${\mathrm{SrTiO}}_{3}$. \emph{Phys. Rev. Research} \textbf{2020}, \emph{2},
  023313\relax
\mciteBstWouldAddEndPuncttrue
\mciteSetBstMidEndSepPunct{\mcitedefaultmidpunct}
{\mcitedefaultendpunct}{\mcitedefaultseppunct}\relax
\EndOfBibitem
\end{mcitethebibliography}

\end{document}